\documentclass[aps,prl,showpacs,notitlepage,singlecolumn,superscriptaddress,11pt,floatfix,]{revtex4-1}
\usepackage{amsmath}
\usepackage{amssymb}
\usepackage{amsfonts}
\usepackage{stmaryrd}
\usepackage{wasysym}
\usepackage{graphicx}
\usepackage{multirow}
\usepackage{tabularx}
\usepackage{textcomp}
\usepackage{subfigure}
\usepackage{url}
\usepackage[colorlinks=true,citecolor=blue,urlcolor=blue]{hyperref}
\usepackage{romannum}
\usepackage{mathtools}
\usepackage[utf8]{inputenc}
\usepackage{braket}
\usepackage{amsmath}
\usepackage{xcolor}
\usepackage{colortbl}
\usepackage{color,soul} 
\usepackage{bm}
\usepackage[normalem]{ulem}
\usepackage[utf8]{inputenc}
\usepackage{array}
\usepackage{makecell}
\usepackage{multirow}
\usepackage{enumitem}  
\hypersetup{colorlinks=true, urlcolor=blue, citecolor=cyan, pdfborder={0 0 0}}

\usepackage{fancyhdr}
\usepackage{pdfpages} 
\usepackage{pgffor} 

\makeatletter
\AtBeginDocument{\let\LS@rot\@undefined}
\makeatother
\def\supplementfilename{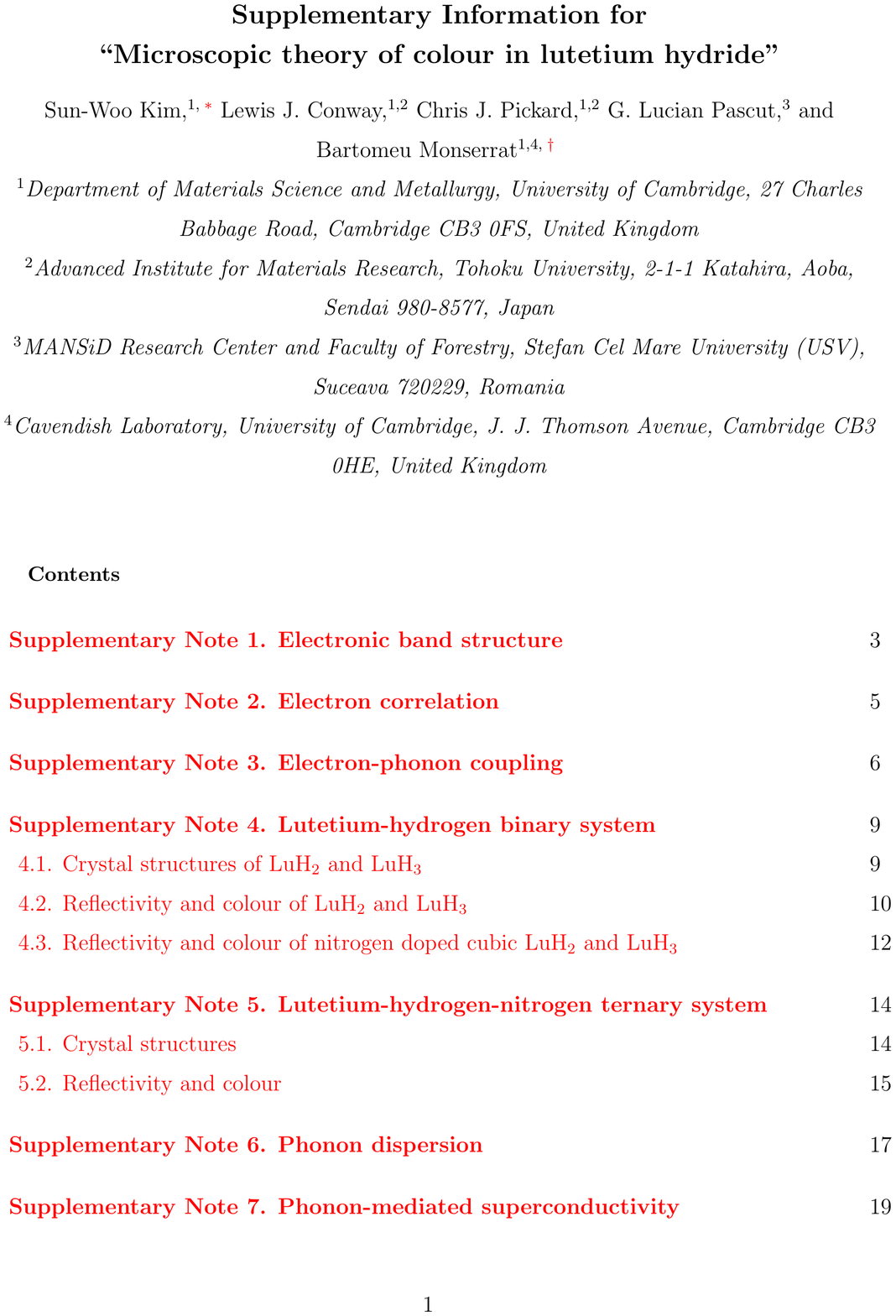}

\pdfximage{\supplementfilename}
\def\numbersupplementpages{\the\pdflastximagepages}

\newif\ifarXiv
\arXivtrue 

\begin{document}
\pagenumbering{arabic}

\title{Microscopic theory of colour in lutetium hydride}
\maketitle
\begin{center}
{Sun-Woo Kim,$^{1,\;\textcolor{red}{*}}$ Lewis J. Conway,$^{1,2}$ Chris J.  Pickard,$^{1,2}$\\ G. Lucian Pascut,$^{3}$ and Bartomeu Monserrat$^{1,4,\;\textcolor{red}{\dagger}}$
}

\emph{$^{1}$Department of Materials Science and Metallurgy, University of Cambridge,\\ 27 Charles Babbage Road, Cambridge CB3 0FS, United Kingdom}\\
\emph{$^{2}$Advanced Institute for Materials Research, Tohoku University,\\ 2-1-1 Katahira, Aoba, Sendai 980-8577, Japan}\\
\emph{$^{3}$MANSiD Research Center and Faculty of Forestry,\\ Stefan Cel Mare University (USV),
Suceava 720229, Romania}\\
\emph{$^{4}$Cavendish Laboratory, University of Cambridge,\\ J. J. Thomson Avenue, Cambridge CB3 0HE, United Kingdom}\\
$^{\textcolor{red}{*}}$ \textup{email: \href{mailto:swk38@cam.ac.uk}{swk38@cam.ac.uk}}\\
$^{\textcolor{red}{\dagger}}$ \textup{email: \href{mailto:bm418@cam.ac.uk}{bm418@cam.ac.uk}}

\end{center}

\section{Abstract}
Nitrogen-doped lutetium hydride has recently been proposed as a near-ambient-conditions superconductor. Interestingly, the sample transforms from blue to pink to red as a function of pressure, but only the pink phase is claimed to be superconducting. 
Subsequent experimental studies have failed to reproduce the superconductivity, but have observed pressure-driven colour changes including blue, pink, red, violet, and orange. However, discrepancies exist among these experiments regarding the sequence and pressure at which these colour changes occur. 
Given the claimed relationship between colour and superconductivity, understanding colour changes in nitrogen-doped lutetium hydride may hold the key to clarifying the possible superconductivity in this compound. 
Here, we present a full microscopic theory of colour in lutetium hydride, revealing that hydrogen-deficient LuH$_2$ is the only phase which exhibits colour changes under pressure consistent with experimental reports, with a sequence blue-violet-pink-red-orange. The concentration of hydrogen vacancies controls the precise sequence and pressure of colour changes, rationalising seemingly contradictory experiments.
Nitrogen doping also modifies the colour of LuH$_2$ but it plays a secondary role compared to hydrogen vacancies. Therefore, we propose hydrogen-deficient LuH$_2$ as the key phase for exploring the superconductivity claim in the lutetium-hydrogen system. Finally, we find no phonon-mediated superconductivity near room temperature in the pink phase. 

\section{Introduction}

The proposal by Ashcroft that hydrogen-rich compounds could host high temperature phonon-mediated superconductivity under pressure\,\cite{hydrogen_superconductivity,ashcroft_superconducting_hydrides} has stimulated a profusion of theoretical proposals for high pressure superconducting hydrides\,\cite{H3S-duan,LaH10-theory-hemley1,LaH10-theory-ma,LaH10-theory-hemley2,pickard-hydride-review,pickard-sc-high-throughput,pickett-hydride-review} and the subsequent experimental discovery of some of these\,\cite{H3S-eremets,LaH10-hemley,LaH10-eremets}. This new class of hydride superconductors has re-ignited the search for superconductivity at ambient conditions, and Dasenbrock-Gammon and co-workers have recently reported superconductivity in nitrogen-doped lutetium hydride with a maximum critical temperature of $294$\,K at a moderate pressure of $10$\,kbar\,\cite{Lu-H-N_superconductivity_nature}. Interestingly, superconductivity is reported to coincide with drastic colour changes in the reflectivity of the sample: increasing pressure transforms a non-superconducting blue metal to a superconducting pink metal at $3$\,kbar, and a further transformation to a non-superconducting red metal above $30$\,kbar. 

This remarkable report has sparked a growing number of experimental\,\cite{exp-colour-change-CPL,exp-reflectivity-scibul,exp-colour-res-mag-wen1,exp-colour-res-mag-wen2,exp-colour-pressure-medium-arxiv,exp-lu-h-n-no-sc-arxiv,exp-elec-mag-arxiv,exp-grain-size-arxiv,exp-luh3-phase-transition-arxiv,luh-sc-high-pressure,salke2023evidence} and theoretical\,\cite{xie-structure-prediction-arxiv,structure-prediction-binary-arxiv,duan-structure-prediction-arxiv,zurek-structure-prediction-arxiv,pickard-structure-prediction-arxiv,ndoping-pressure-luh3-arxiv} investigations, most of which have so far failed at reproducing or explaining near-ambient superconductivity. On the experimental front, most measurements of resistivity and magnetic susceptibility find no superconductivity near ambient conditions, 
with the exception of a recent work in which resistivity changes consistent with high temperature superconductivity are reported\,\cite{salke2023evidence}. 
Puzzlingly, multiple studies report pressure-driven colour changes, but these include a wide range of seemingly incompatible colour sequences and pressure conditions: blue-to-pink at $3$\,kbar and pink-to-red at $30$\,kbar in the original report\,\cite{Lu-H-N_superconductivity_nature}; blue-to-pink at $22$\,kbar and pink-to-red at $40$\,kbar\,\cite{exp-colour-change-CPL}; blue-to-violet upon contact with a diamond culet, violet-to-red at $30$\,kbar and red-to-orange at $120$\,kbar\,\cite{exp-reflectivity-scibul}; blue-to-violet at $94$\,kbar\,\cite{exp-reflectivity-scibul}; blue-to-violet at $120$\,kbar, violet-to-pink-to-red gradually between $160$\,kbar and $350$\,kbar and red persisting up to $420$\,kbar\,\cite{exp-colour-res-mag-wen1,exp-colour-res-mag-wen2}; blue-to-violet-to-pink-to-red with transition pressures differing by up to $60$\,kbar depending on the pressure medium used\,\cite{exp-colour-pressure-medium-arxiv}; and persistent blue colour up to $65$\,kbar\,\cite{exp-lu-h-n-no-sc-arxiv}. Growing evidence suggests that the colour changes are significantly affected by the initial compression procedure\,\cite{exp-reflectivity-scibul} and by the pressure medium used in the diamond anvil cell\,\cite{exp-colour-pressure-medium-arxiv}.

On the theoretical front there have been multiple reports of structure searches in the Lu-H binary and the Lu-H-N ternary systems\,\cite{xie-structure-prediction-arxiv,structure-prediction-binary-arxiv,duan-structure-prediction-arxiv,zurek-structure-prediction-arxiv,pickard-structure-prediction-arxiv}. Most studies only report metastable ternary structures, but Ferreira and co-workers report a ternary Lu$_4$N$_2$H$_5$ stable structure\,\cite{pickard-structure-prediction-arxiv}. The roles of pressure and nitrogen doping\,\cite{ndoping-pressure-luh3-arxiv} and of quantum and thermal ionic vibrations\,\cite{anharmonic-luh3-arxiv} in stabilising the cubic LuH$_3$ structure have also been studied. None of the predicted stable or metastable structures are found to be phonon-mediated superconductors near room temperature. 
 
Given the claimed association between superconductivity and colour changes in the original superconductivity report, understanding colour changes in nitrogen-doped lutetium hydride holds the key to clarifying the possible superconductivity in this compound. However, experimental reports provide an inconsistent picture regarding pressure-driven colour changes, and there are no theoretical studies yet. In this work, we provide a full microscopic theory of colour in lutetium hydride.

\section{Results}

\subsection{LuH$_2$ under ambient conditions}

\begin{figure}[ht]
 \centering
 \includegraphics[width=1\textwidth]{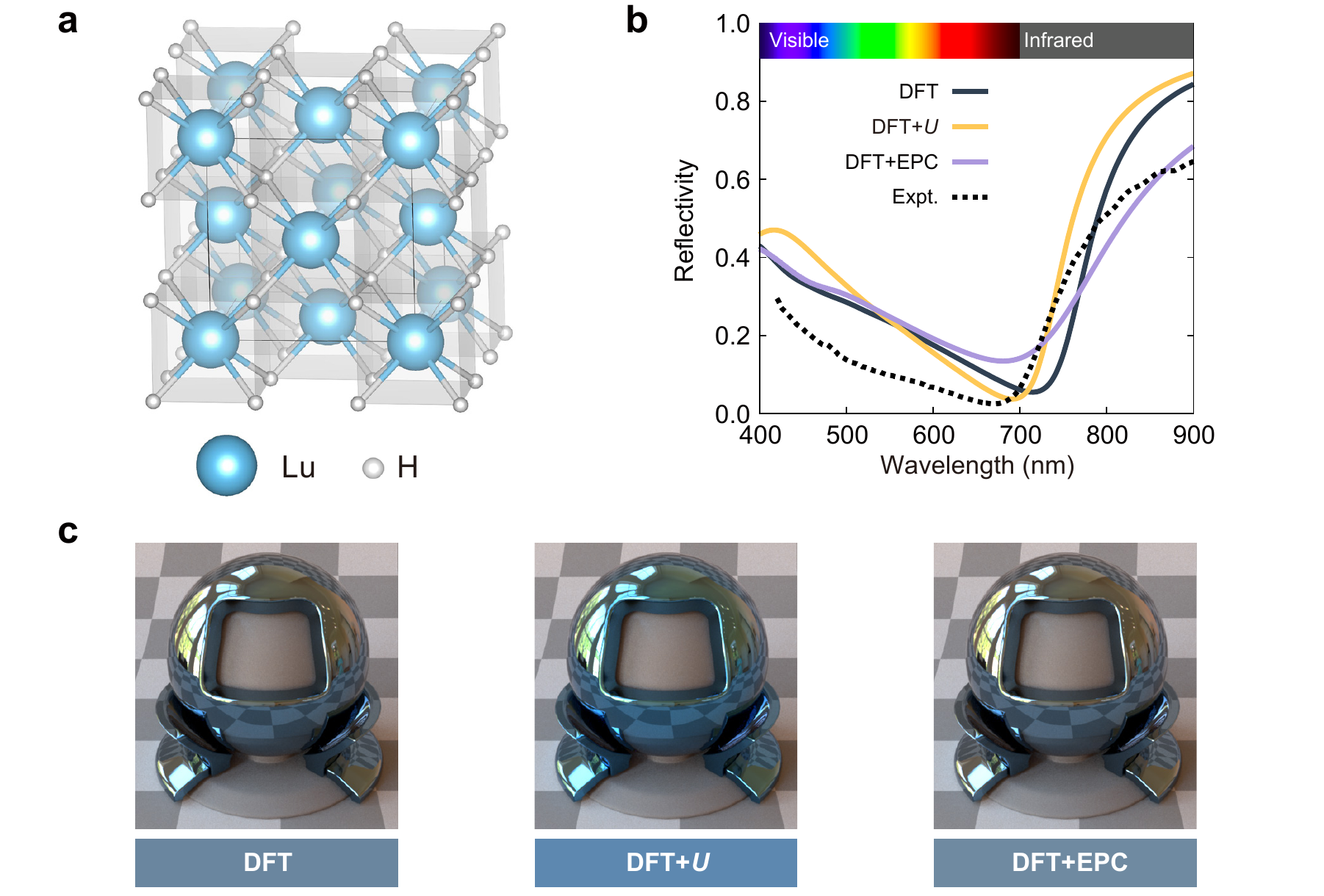} 
  \caption{ \textbf{Structure, reflectivity, and colour of LuH$_2$.}
  \textbf{a.} Crystal structure of $Fm\overline{3}m$ LuH$_2$. \textbf{b.} Reflectivity of LuH$_2$ calculated using semilocal density functional theory (DFT), DFT corrected with a Hubbard $U$ term (DFT$+U$), and DFT including electron-phonon coupling (DFT$+$EPC). The experimental reflectivity is taken from Ref.\,\cite{exp-reflectivity-scibul}. \textbf{c.} Colour and photorealistic rendering of LuH$_2$ calculated using DFT, DFT$+U$, and DFT$+$EPC. The photorealistic rendering is shown as LuH$_2$ surrounding a grey ball with an opening in the centre.
  }
  \label{fig:luh2-blue}%
\end{figure}

Lutetium hydride under ambient conditions crystallises in the LuH$_2$ stoichiometry with cubic space group $Fm\overline{3}m$. As shown in Fig.\,\ref{fig:luh2-blue}, LuH$_2$ adopts the fluorite structure with the lutetium atoms occupying the sites of an fcc lattice, and the hydrogen atoms occupying the tetrahedral interstitial sites. 

LuH$_2$ is a metal whose reflectivity endows it with a blue appearance. We demonstrate the validity of our computational approach by reporting the calculated colour of LuH$_2$ at ambient conditions in Fig.\,\ref{fig:luh2-blue}. We show results using three distinct computational models to explore the potential role of electron correlation due to the presence of lutetium $5d$ electrons and the potential role of strong electron-phonon coupling due to the presence of hydrogen.   

The first model we consider uses semilocal density functional theory (DFT) in the generalised gradient approximation, labelled DFT in Fig.\,\ref{fig:luh2-blue}. This model provides a basic description of the electronic structure without a detailed treatment of electron correlation and without the inclusion of electron-phonon effects. The calculated reflectivity is large in the infrared region above $800$\,nm, is strongly suppressed in the red part of the visible spectrum with a calculated minimum at $710$\,nm, and increases gradually towards the blue part of the visible spectrum. The overall shape of the reflectivity is consistent with that observed experimentally\,\cite{exp-reflectivity-scibul} and directly leads to the blue colour of LuH$_2$.

Lutetium has an electronic configuration with a partially filled $5d$ shell, which suggests that electronic correlation beyond that captured by standard DFT may contribute to the electronic properties of LuH$_2$. To explore the possible role of electron correlation, we repeat our calculations using DFT corrected with a Hubbard $U$ term, labelled as DFT$+U$ in Fig.\,\ref{fig:luh2-blue}, which captures static correlation. The reflectivity curve has a similar shape to that obtained at the DFT level, but the minimum of the reflectivity has a lower value and occurs at a slightly shorter wavelength of $690$\,nm. Combined with a slighly larger reflectivity in the blue part of the visible spectrum, we obtain a slightly brighter blue colour for LuH$_2$ using the DFT$+U$ model. These results indicate that static electron correlation arising from lutetium only plays a minor role in LuH$_2$. We have also performed dynamical mean field theory calculations that capture dynamical correlation and also find that they can be neglected. We rationalise these results by noting that $5d$ orbitals have a large electron bandwidth spanning multiple eV (see Supplementary Fig.\,S1) which implies that the spatial extent of the orbitals is large and the corresponding local correlations weak. 

Hydrogen is the lightest of all elements, and as such it exhibits significant nuclear motion, even at zero temperature, due to quantum zero-point effects. This significant nuclear motion can lead to strong electron-phonon coupling, and this is indeed the prime motivation behind the proposal that hydrogen-rich compounds could be high temperature phonon-mediated superconductors. To explore the possible role of electron-phonon interactions in LuH$_2$, we repeat our calculations including contributions from both zero-point quantum nuclear motion at $0$\,K and thermal nuclear motion at finite temperature, labelled DFT$+$EPC in Fig.\,\ref{fig:luh2-blue}. The reflectivity curve has a similar shape to those obtained with DFT and DFT$+U$, but exhibits a lower value in the infrared region and a larger value in the red region. We again obtain a blue colour, indicating that electron-phonon coupling does not significantly modify the reflectivity of LuH$_2$. 

Overall, we find that electronic correlation and electron-phonon interactions make a small contribution, and that the main features of the reflectivity curve and the resulting blue colour of LuH$_2$ are correctly captured by semilocal DFT. Therefore, our subsequent discussion neglects electron correlation and electron-phonon interactions, but further details about these contributions are included in the Supplementary Information.

\subsection{Lutetium hydride colour changes under pressure}

\begin{figure}[ht]
 \centering
 \includegraphics[width=1\textwidth]{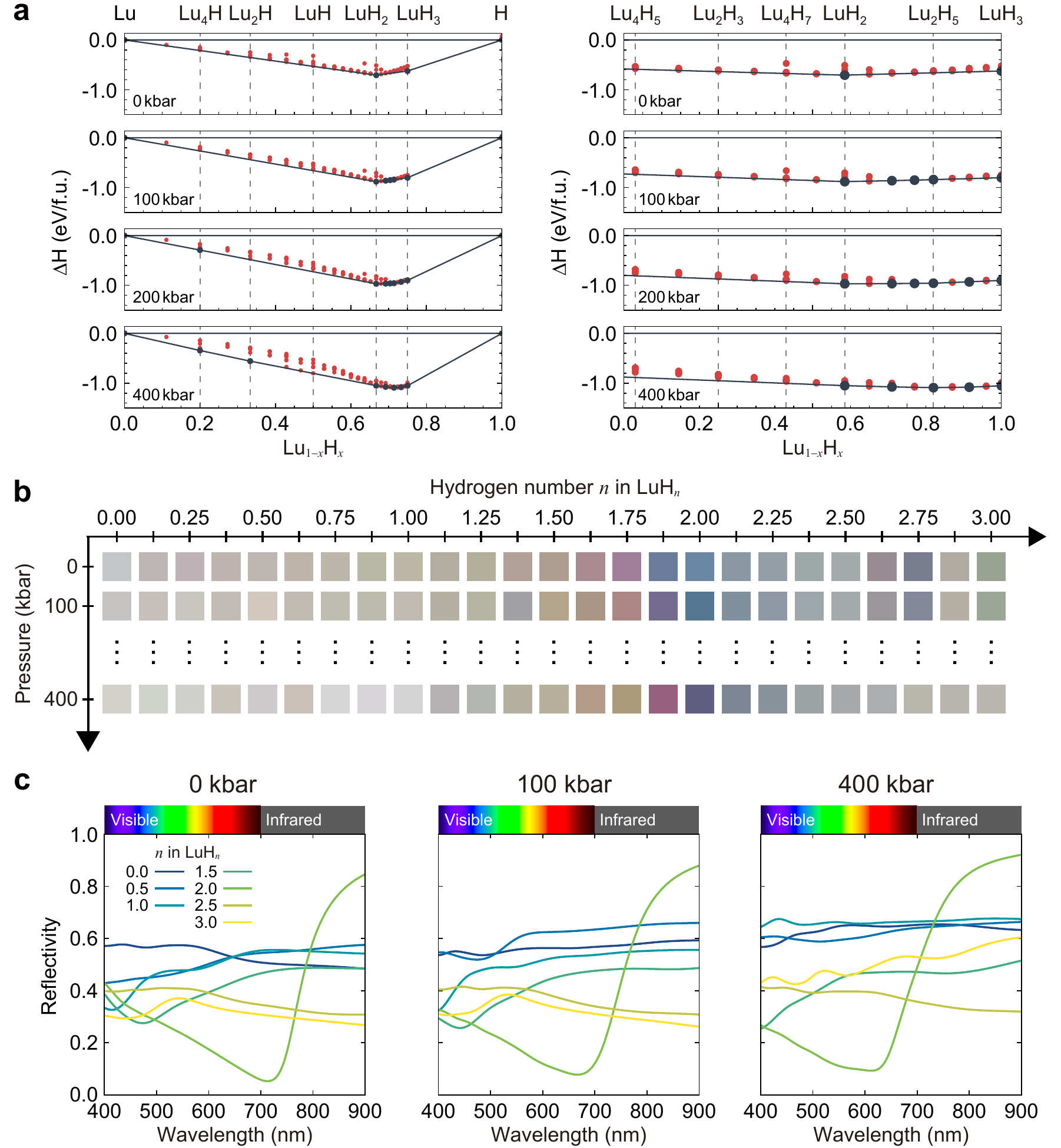} 
  \caption{
  \textbf{Pressure dependence of the convex hull diagram, colour, and reflectivity of the lutetium-hydrogen binary system.}
  \textbf{a.} Convex hull diagrams as a function of pressure for compositions in the range LuH$_0$ to LuH$_3$. Dark blue circles indicate thermodynamically stable structures and red circles indicate metastable structures. The dashed vertical lines for selected stoichiometries are shown for guidance only. \textbf{b.} Colour of LuH$_n$ compounds as a function of pressure. \textbf{c.} Reflectivity of LuH$_n$ compounds as a function of pressure.
  }
  \label{fig:lu-h-colour-changes}%
\end{figure}

To build a microscopic theory of colour in lutetium hydride, we first explore the pressure-driven colour changes in the lutetium-hydrogen system. We have performed extensive structure searches for stoichiometries ranging from LuH$_0$ to LuH$_3$ at multiple pressures. The results are summarised in the convex hull diagrams depicted in Fig.\,\ref{fig:lu-h-colour-changes}a. At $0$\,kbar, the only thermodynamically stable structures are LuH$_2$ ($Fm\overline{3}m$ space group) and LuH$_3$ ($P\overline{3}c1$ space group; not cubic). There are multiple metastable structures in the entire composition space from LuH$_0$ to LuH$_3$ that are within $60$\,meV/atom of the convex hull. Increasing pressure leads to multiple additional stable structures with stoichiometries intermediate between LuH$_2$ and LuH$_3$. We highlight that substoichiometric LuH$_{2-\delta}$ structures are close to the convex hull at both $0$ and $400$\,kbar (within $11$ and $42$\,meV/atom, respectively, up to $\delta=0.25$). We have also checked their dynamical stability (see Supplementary Fig.\,S14), so we expect that they can be accessible experimentally.

We show the calculated colour as a function of pressure for the most stable structure at each composition between LuH$_0$ and LuH$_3$ in Fig.\,\ref{fig:lu-h-colour-changes}b. Consistently with experimental observations, we find that pure lutetium has a silvery white colour\,\cite{exp-elec-mag-arxiv} (see also Supplementary Fig.\,S17) and, as described above in Fig.\,\ref{fig:luh2-blue}, LuH$_2$ has a blue colour. Across the entire composition space, the only compositions that exhibit a blue colour at ambient pressure occur for stoichiometries close to LuH$_2$. Similarly, the only stoichiometries that exhibit a violet colour at high pressure are those close to LuH$_2$. Specifically, we find this trend is present for substoichiometric LuH$_{2-\delta}$, but not present for suprastoichimoetric LuH$_{2+\delta}$. We also note that LuH$_3$ has a grey-green colour at ambient pressure that becomes grey with increasing pressure.

Figure\,\ref{fig:lu-h-colour-changes}c shows reflectivity curves for selected stoichiometries in the range LuH$_0$ to LuH$_3$ at multiple pressures. We note that only the reflectivity of LuH$_2$ has a minimum in the red part of the visible spectrum leading to an overall blue colour, as already discussed in Fig.\,\ref{fig:luh2-blue} above. The reflectivities of all other compositions show relatively flat curves across the visible spectrum, which lead to colours with various tones of grey.

Overall, the results depicted in Fig.\,\ref{fig:lu-h-colour-changes} show that the only compounds in the lutetium-hydrogen binary space that are blue at ambient conditions and violet at high pressure have stoichiometries close to LuH$_2$ with a moderate amount of hydrogen vacancies.
This conclusion still holds when structures in the ternary lutetium-hydrogen-nitrogen system are considered. 
To demonstrate this, we perform extensive crystal structure searches in the full Lu-H-N ternary space (see Supplementary Fig.\,S12), and note that our structure searches are the only ones of all those published that identify a stable ternary compound at ambient pressure\,\cite{pickard-structure-prediction-arxiv}. We have calculated the colour of this stable compound and also the colour of multiple other ternary compounds that are not on the convex hull but whose simulated X-ray diffraction data is consistent with experimental reports. We find that none of these structures
exhibit colours that are consistent with experiment (see Supplementary Fig.\,S13).

These observations allow us to conclude that the colour changes in the lutetium-hydrogen-nitrogen system are dominated by the LuH$_2$ stoichiometry. In particular, we discard the LuH$_3$ composition proposed by Dasenbrock-Gammon and co-workers to explain high temperature superconductivity\,\cite{Lu-H-N_superconductivity_nature} as this structure has a grey-green colour at all pressures. 
LuH$_2$ has also been identified as the relevant stoichiometry by comparing the calculated equation of state\,\cite{zurek-structure-prediction-arxiv} and X-ray diffraction patterns\,\cite{zurek-structure-prediction-arxiv,structure-prediction-binary-arxiv,pickard-structure-prediction-arxiv,xie-structure-prediction-arxiv} to experiment, and we also note a recent experimental work that uses LuH$_2$-based samples and that has successfully reproduced the Raman spectrum and colour sequence with pressure reported in the original work\,\cite{exp-colour-res-mag-wen1}.

\subsection{Hydrogen deficient LuH$_{2-\delta}$}

The calculated colour changes from blue to violet in LuH$_2$ are also observed in multiple experiments\,\cite{exp-reflectivity-scibul,exp-colour-res-mag-wen1,exp-colour-res-mag-wen2,exp-colour-pressure-medium-arxiv}, but they occur at different pressures in different experiments, ranging from $0$ to at least $190$\,kbar, but possibly higher as some experiments only observe a blue phase. Furthermore, some experimental observations reveal additional colour changes with increasing pressure, which include pink\,\cite{Lu-H-N_superconductivity_nature,exp-colour-change-CPL,exp-colour-pressure-medium-arxiv}, red\,\cite{Lu-H-N_superconductivity_nature,exp-colour-change-CPL,exp-reflectivity-scibul,exp-colour-pressure-medium-arxiv}, and orange\,\cite{exp-reflectivity-scibul}. The pink colour is particularly important as it is associated with the superconducting phase in the original report\,\cite{Lu-H-N_superconductivity_nature}. Based on these observations, we next explore the pressure-driven colour changes of substoichiometric LuH$_2$ in more detail. 

\begin{figure}[ht]
 \centering
 \includegraphics[width=0.9\textwidth]{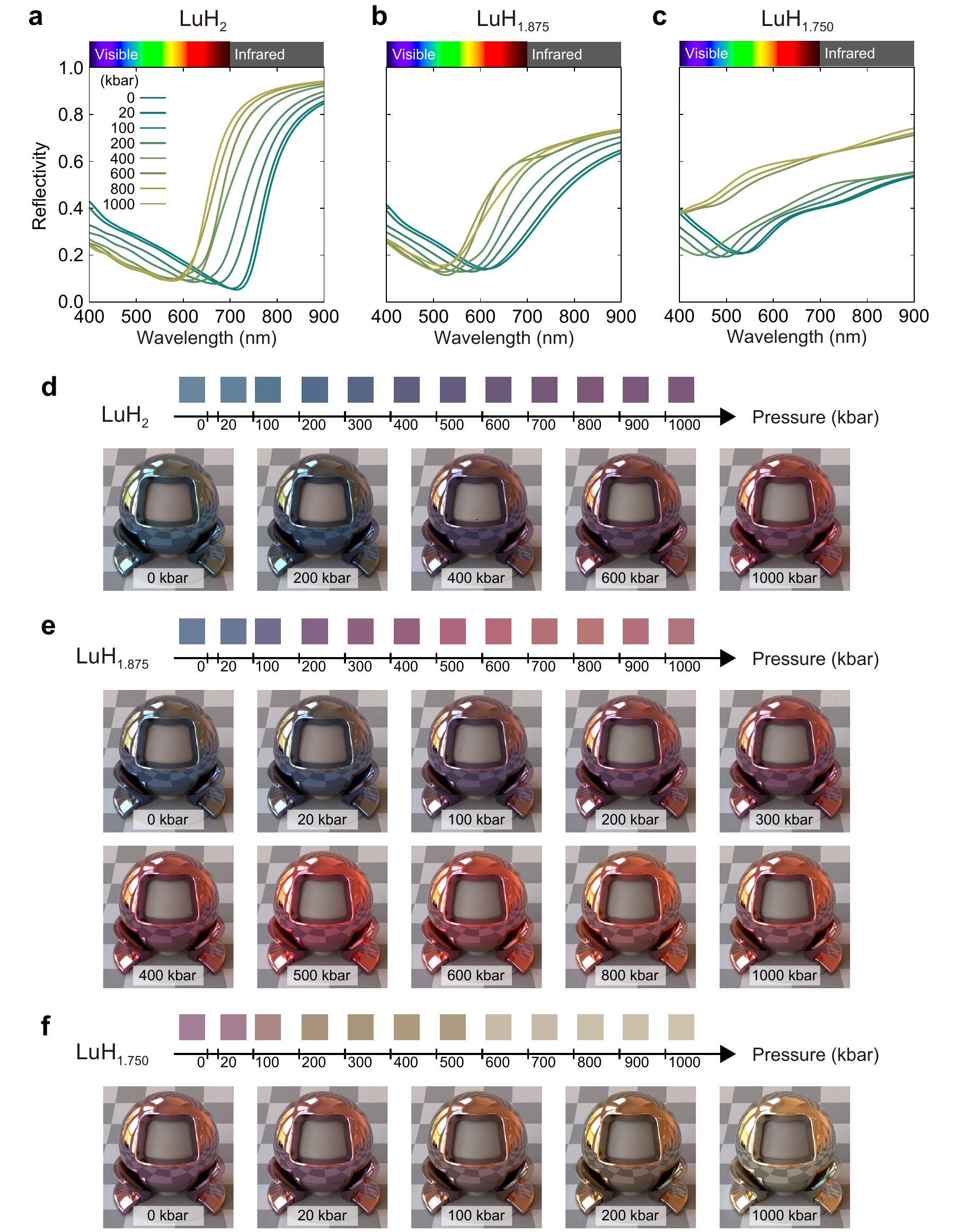} 
  \caption{
  \textbf{Pressure dependence of the reflectivity and colour of pure and hydrogen-deficient lutetium dihydrides.}
  \textbf{a-c.} Reflectivity as a function of pressure (in kbar) for \textbf{a} LuH$_{2}$, \textbf{b} LuH$_{1.875}$ and \textbf{c} LuH$_{1.750}$. \textbf{d-f.} Colour and photorealistic rendering of \textbf{d} LuH$_{2}$, \textbf{e} LuH$_{1.875}$ and \textbf{f} LuH$_{1.750}$ as a function of pressure.   }
  \label{fig:h-deficiency}%
\end{figure}

We show the pressure evolution of the reflectivity and colour of LuH$_2$ and hydrogen-deficient LuH$_{1.875}$ and LuH$_{1.750}$ in Fig.\,\ref{fig:h-deficiency}. At ambient conditions LuH$_2$ has the reflectivity described in Fig.\,\ref{fig:luh2-blue} and repeated in Fig.\,\ref{fig:h-deficiency} with a minimum in the red part of the spectrum that leads to an overall blue colour. With increasing pressure, the reflectivity minimum shifts towards shorter wavelengths, and the reflectivity from the red part of the spectrum increases, in agreement with experiment\,\cite{exp-reflectivity-scibul}. This leads to a gradual colour change from blue to violet with increasing pressure. LuH$_2$ undergoes a structural phase transition at $732$\,kbar to a phase of space group $P4/nmm$, which has a grey colour with an orange-red hue (reflectivity and colour shown in Supplementary Fig.\,S7).

At ambient conditions, LuH$_{1.875}$ exhibits a reflectivity with a shape similar to that of LuH$_2$ but with the minimum occurring at somewhat shorter wavelenghts of $600$\,nm. The resulting colour is still blue. Similar to LuH$_2$, increasing pressure leads to an overall shift of the reflectivity minimum to shorter wavelengths and to an increase in the reflectivity in the red part of the spectrum. As a result we observe a blue-to-violet colour change at a pressure of about $100$\,kbar, significantly lower than the corresponding colour change in pure LuH$_2$ and in the experimentally observed pressure range. Increasing pressure further leads to a gradual transition to pink (peaking at about $300$\,kbar), followed by red (peaking at about $500$\,kbar) and tending towards orange approaching $1000$\,kbar. Therefore, hydrogen-deficient LuH$_2$ exhibits a sequence of colour changes that includes all colours reported experimentally.

The sequence and pressure of colour changes in hydrogen-deficient LuH$_2$ is strongly dependent on the concentration of hydrogen vacancies. Figure\,\,\ref{fig:h-deficiency} also depicts the pressure evolution of the reflectivity and colour changes of LuH$_{1.750}$ with a higher concentration of hydrogen vacancies. In this case, the reflectivity minimum occurs at a wavelength of $550$\,nm at ambient conditions, giving a pink colour. Increasing pressure suppresses the reflectivity in the blue region, turning the colour from pink towards orange at lower pressures than those necessary for LuH$_{1.875}$.

These results suggest that the seemingly contradictory experimental observations of colour changes in lutetium hydride are likely due to varying hydrogen vacancy concentrations in LuH$_2$. In particular, Dasenbrock-Gammon and co-workers observe a pink phase starting with a pressure of $3$\,kbar\,\cite{Lu-H-N_superconductivity_nature}, significantly lower than the pressure reported in multiple subsequent experiments. Our results suggest that this is due to a higher concentration of hydrogen vacancies in the original work compared to subsequent studies. 

The concentration of nitrogen dopants also induces colour changes in LuH$_2$ (see Supplementary Fig.\,S10), but the colour changes driven by nitrogen doping only play a secondary role compared to hydrogen vacancies. We have also tested the role that hydrogen vacancies and nitrogen doping have on the reflectivity and colour of cubic $Fm\overline{3}m$ LuH$_3$ (see Supplementary Fig.\,S11), as this phase has been tentatively identified as the parent phase responsible for the superconductivity claim\,\cite{Lu-H-N_superconductivity_nature}. Our results show that the LuH$_3$ phase retains a grey colour under nitrogen doping, further discarding it as a relevant phase.

\subsection{Absence of phonon-mediated superconductivity in the lutetium-hydrogen-nitrogen system}

Most experimental reports since the original announcement of near-ambient conditions superconductivity in the lutetium-hydrogen-nitrogen system have been unable to confirm this claim. Similarly, no calculation of stable and metastable phases in the lutetium-hydrogen-nitrogen system has predicted a high superconducting critical temperature within a phonon-mediated framework.

\begin{center}
\begin{table}[b]
\caption{
\textbf{Calculated superconducting properties of hydrogen-deficient LuH$_2$ at various pressures.} $\lambda$ is the total electron-phonon coupling parameter computed from the
Eliashberg function, $\omega_{\text{log}}$ is the logarithmic averaged frequency,  $T_c^{\mathrm{AD}}$ is the superconducting critical temperature estimated from the semiempirical Allen-Dynes formula\,\cite{TC_AD_formula}, and $T_c^{\mathrm{E}}$ is the superconducting critical temperature estimated from the isotropic Eliashberg equation. A standard value of $\mu^*=0.125$ is used for the Morel-Anderson Coulomb pseudopotential. 
}
\renewcommand{\arraystretch}{1.2}
\setlength{\tabcolsep}{12pt} 
\begin{tabular}{|c|c|c|c|c|c|}
\hline \hline
Structure & Pressure & $\lambda$ & $\omega_{\text{log}}$ & $T_c^{\mathrm{AD}}$  & $T_c^{\mathrm{E}}$ \\
\hline
\multirow{3}{*}{LuH$_{1.875}$}
 & 20\,kbar   & 0.333 & 257.516 & 0.107 & 0.190 \\
\cline{2-6}
 & 100\,kbar & 0.326 & 257.997 & 0.085 & 0.161  \\
\cline{2-6}
 & 400\,kbar & 0.360 & 243.566 & 0.220 & 0.328\\
\hline
\multirow{3}{*}{LuH$_{1.750}$}
 & 20\,kbar   & 0.277 & 231.482 & 0.008 & 0.035   \\
\cline{2-6}
 & 100\,kbar & 0.276 & 224.351 & 0.007 &  0.032  \\
\cline{2-6}
 & 400\,kbar & 0.299 & 211.951 & 0.023  & 0.062  \\
\hline \hline
\end{tabular}
\end{table}
\end{center}

Our results suggest that hydrogen-deficient LuH$_2$ is responsible for the colour changes observed experimentally. Given the claim by Dasenbrock-Gammon and co-workers that it is the pink phase that is a room temperature superconductor, we calculate the superconducting critical temperature of LuH$_{1.875}$ under pressure, which exhibits the pink phase. We find no room-temperature superconductivity, with a calculated critical temperature of the order of $0.1$\,K (Table I; see also details in Supplementary Note 7).

\section{Discussion}

Dasenbrock-Gammon and co-workers report superconductivity in nitrogen-doped lutetium hydride over the pressure range $3$-$30$\,kbar\,\cite{Lu-H-N_superconductivity_nature}. Importantly, the claimed pressure-driven transition to and from the superconducting phase occurs simultaneously with drastic colour changes in the sample, which is pink in the claimed superconducting phase, compared to blue below $3$\,kbar and red above $30$\,kbar. Additionally, they attribute the claimed superconductivity to a cubic LuH$_3$ phase with some unknown concentration of nitrogen dopants and some unknown concentration of hydrogen vacancies.

Our calculations show that the cubic LuH$_3$ phase is not consistent with the colour changes observed experimentally. Additionally, our results suggest that the only phase that is consistent with the observed colour changes is hydrogen-deficient cubic LuH$_2$, and that the concentration of hydrogen vacancies and nitrogen dopants controls the colour at each pressure. Finally, we also show that hydrogen-deficient LuH$_2$ is unlikely to be a high temperature phonon-mediated superconductor.

Our work presents a compelling demonstration of how the first principles prediction of the colour of a material can be exploited to effectively identify the microscopic characteristics of the corresponding experimental samples. Given that colour is readily accessible experimentally, while other structural characterisation techniques can be challenging to implement (particularly under pressure), our work provides a promising new avenue for identifying composition and structure of complex samples. It would be interesting to further explore the applicability of this method to study the colour of other compounds, including strongly correlated materials\,\cite{Iron_SC_1,Iron_SC_2} and magnetic materials\,\cite{AFM_1,AFM_2}.

\section{Methods}

{\em Electronic structure calculations. -}
We perform density functional theory (DFT) calculations using the Vienna $ab$ $initio$ simulation package ({\sc vasp})\,\cite{VASP1,VASP2} implementing the projector-augmented wave (PAW) method~\cite{PAW}. We treat the $4f$ states of lutetium as valence by employing PAW pseudopotentials with 25 valence electrons ($4f^{14} 5s^2 5p^6 5d^1 6s^2$). 
For the exchange-correlation energy, we use the generalized-gradient approximation functional of Perdew-Burke-Ernzerhof modified for solids (PBEsol)\,\cite{PBEsol}. Converged results are obtained with a kinetic energy cutoff for the plane wave basis of $400$\,eV and a $\mathbf{k}$-point grid of size $40\times40\times40$ for the LuH$_2$ primitive cell and commensurate grids for other cell sizes and shapes (see convergence tests in Supplementary Fig.\,S20). The geometry of the structures is optimised until all forces are below $0.01$\,eV/\AA\@ and the pressure is below $1$\,kbar. 
We also perform select calculations using DFT corrected with a Hubbard $U$ term, and for these we use a value of $U=3$\,eV. We also perform select calculations using dynamical mean field theory with the e{\sc dmft} code\,\cite{edmft-1,edmft-2}, which implements density functional theory with embedded dynamical mean field theory (DFT+eDFMT). For the DFT part we have used the {\sc wien2k} code\,\cite{wien2k}. 

{\em Reflectivity and colour. -} 
Our reflectivity calculations follow the methodology described in Ref.\,\cite{colour-marzari}. We calculate the complex dielectric function within the independent-particle approximation as implemented in {\sc vasp}. In the optical limit ($\mathbf{q}\rightarrow 0$), the dielectric function $\varepsilon(\mathbf{q},\omega)$ is given by the sum of an intraband Drude-like term $\varepsilon^{\mathrm{intra}}(\mathbf{q},\omega)$ due to the electrons at the Fermi surface and an interband term $\varepsilon^{\mathrm{inter}}(\mathbf{q},\omega)$ describing vertical transitions between valence and conduction bands. The explicit form of each term is given by\,\cite{dielectric_fct1,dielectric_fct2}:
 \begin{equation}
    \varepsilon^{\mathrm{intra}}(\mathbf{q},\omega)=-\frac{\omega^2_D(\mathbf{\hat{q})}}{\omega(\omega+i\gamma)},
 \end{equation}
 where the independent particle approximation Drude plasma frequency is 
 \begin{equation}
   \omega^2_D(\mathbf{\hat{q})}=\frac{4\pi}{V}\sum_{\mathbf{k}}\sum_n\left|\bra{\psi_{n\textbf{k}}}\mathbf{\hat{q}}\cdot\mathbf{v}\ket{\psi_{n\mathbf{k}}}\right|^2 \left(-\frac{\partial f_{n\mathbf{k}}}{\partial E_{n\mathbf{k}}}\right)  ,
 \end{equation}
and
 \begin{equation}
    \varepsilon^{\mathrm{inter}}(\mathbf{q},\omega)=1-\frac{4\pi}{V}\sum_{\mathbf{k}}\sum_{n;n\neq n'}\sum_{n'}\frac{\left|\bra{\psi_{n\textbf{k}}}\mathbf{\hat{q}}\cdot\mathbf{v}\ket{\psi_{n'\mathbf{k}}}\right|^2}{(E_{n\mathbf{k}}-E_{n'\mathbf{k}})^2} \frac{f_{n\mathbf{k}}-f_{n'\mathbf{k}}}{\omega+E_{n\mathbf{k}}-E_{n'\mathbf{k}}+i\eta}.
 \end{equation}
In these equations, $V$ is the volume of the system, $|\psi_{n\mathbf{k}}\rangle$ is an electronic state with associated energy $E_{n\mathbf{k}}$ and labelled with quantum numbers $(n,\mathbf{k})$, $\mathbf{v}$ is the velocity operator, and $f_{n\mathbf{k}}$ is the Fermi-Dirac distribution.
We use the empirical broadening parameters $\gamma=\eta=0.1$\,eV.
To obtain the reflectivity, we average the dielectric function and the Drude plasma frequency as 
 \begin{equation}    \varepsilon(\omega)=\frac{\varepsilon(\mathbf{\hat{x}},\omega)+\varepsilon(\mathbf{\hat{y}},\omega)+\varepsilon(\mathbf{\hat{z}},\omega)}{3} \hspace{1cm}\text{and}\hspace{1cm} \omega^2_D=\frac{\omega^2_D(\mathbf{\hat{x}})+\omega^2_D(\mathbf{\hat{y}})+\omega^2_D(\mathbf{\hat{z}})}{3}.
 \end{equation}
Using the relation $\varepsilon(\omega)=[n(\omega)+ik(\omega)]^2$ with the refractive index $n(\omega)$ and the optical extinction coefficient $k(\omega)$, we compute the reflectivity at normal incidence by assuming a vacuum-material interface as 
 \begin{equation}    R(\omega)=\frac{[n(\omega)-1]^2+k(\omega)^2}{[n(\omega)+1]^2+k(\omega)^2}.
 \end{equation}

We follow the method described in Ref.\,\cite{colour-marzari} to obtain the colour from the reflectivity and we assign names to the calculated colours based on the online tool {\sc ArtyClick}\,\cite{artyclick}.
We note that the intensity of the colour can change slightly using different exchange-correlation functionals (see Supplementary Fig.\,S19).

We use the {\sc Mitsuba 3} renderer for the photorealistic rendering\,\cite{mitsuba3}. 
The photorealistic images presented in the main text are rendered by assuming an ideal bulk system with a clean surface. We have also considered the effects of surface roughness on the colour and find no significant changes (see Supplementary Fig.\,S18).
We note that colour perception is subjective and that colour appearance depends on many effects including surface roughness and thickness of the sample. For this reason, we include all calculated dielectric functions as Supplementary Data, which allows readers to independently explore the resulting colours using their own setup.

{\em Electron-phonon coupling calculations. -}
We use the finite displacement method in conjunction with nondiagonal supercells\,\cite{nondiagonal_supercells} to calculate the phonon frequencies $\omega_{\mathbf{q}\nu}$ and eigenvectors $\mathbf{e}_{\mathbf{q}\nu}$ of a phonon mode labelled by wave vector $\mathbf{q}$ and branch $\nu$. The electronic structure parameters are the same as those reported above, and we use a $4\times4\times4$ coarse $\mathbf{q}$-point grid to construct the matrix of force constants. Representative phonon dispersions are reported in Supplementary Note 6. We evaluate the imaginary part of the dielectric function at temperature $T$ renormalized by electron-phonon coupling using the Williams-Lax theory\,\cite{Williams1951,Lax1952}:
 \begin{equation}
     \varepsilon_{2}(\omega ; T)=\frac{1}{\mathcal{Z}} \sum_{\mathbf{s}}\left\langle\Phi_{\mathbf{s}}(\mathbf{u})\left|\varepsilon_{2}(\omega ; \mathbf{u})\right| \Phi_{\mathbf{s}}(\mathbf{u})\right\rangle e^{-E_{\mathbf{s}} / k_{\mathrm{B}} T}, \label{eq:williams-lax}
 \end{equation}
where $\mathcal{Z}$ is the partition function, $|\Phi_{\mathbf{s}}(\mathbf{u})\rangle$ is a harmonic eigenstate $\mathbf{s}$ of energy $E_{\mathbf{s}}$, $\mathbf{u}=\{u_{\mathbf{q}\nu}\}$ is a vector containing all atomic positions expressed in terms of normal mode amplitudes $u_{\mathbf{q}\nu}$, and $k_{\mathrm{B}}$ is Boltzmann's constant. We evaluate Eq.\,(\ref{eq:williams-lax}) by Monte Carlo integration accelerated with thermal lines\,\cite{thermal_lines}: we generate atomic configurations in which the atoms are distributed according to the harmonic nuclear wave function in which every normal mode has an amplitude of $\left(\frac{1}{2\omega_{\mathbf{q}\nu}}\left[1+2n_{\mathrm{B}}(\omega_{\mathbf{q}\nu},T)\right]\right)^{1/2}$, where $n_{\mathrm{B}}(\omega,T)$ is the Bose-Einstein factor. We note that the two terms in the Bose-Einstein factor imply that the electron-phonon renormalised dielectric function includes the effects of both quantum zero-point nuclear vibrations at $T=0$\,K and thermal nuclear vibrations at finite temperature. We build the electron-phonon renormalised reflectivity using the electron-phonon renormalised dielectric function.

For the superconductivity calculations we evaluate the electronic and vibrational properties using {\sc Quantum Espresso}\,\cite{quantum_espresso} together with GBRV ultrasoft pseudopotentials\,\cite{gbrv_pseudos} and the PBE exchange-correlation functional\,\cite{pbe}. We use a plane-wave cutoff of $50$\,Ry, a $4\times4\times4$ $\mathbf{k}$-point grid, and a Gaussian smearing of $0.02$\,Ry for geometry optimisations and electronic structure calculations. 
We use phonon $\mathbf{q}$-grids of size $4\times4\times4$ and dense electron $\mathbf{k}$-grids of size $8\times8\times8$ for the superconductivity calculations. In the calculation of the Eliashberg spectral function,
\begin{equation}
\alpha^2 F(\omega) = \frac{1}{2} \sum_\nu \int_\text{BZ} \frac{d\mathbf{q}}{\Omega_\text{BZ}} \lambda_{\mathbf{q}\nu} \delta{(\omega-\omega_{\mathbf{q}\nu})},
\end{equation}
the integral is calculated by a sum over the $\mathbf{q}$-grids and the Dirac-delta functions are replaced by Gaussians with a width of 0.1\,THz, where $\lambda_{\mathbf{q}\nu}$ is the electron-phonon coupling parameter for vibrational mode $\nu$ at $\mathbf{q}$. These parameters provide converged values of $\lambda$ and $\omega_\text{log}$ derived from moments of $\alpha^2F(\omega)$. We estimate the superconducting critical temperature $T^\text{AD}_c$ using the McMillan-Allen-Dynes formula\,\cite{TC_AD_formula},
\begin{equation}
k_{\mathrm{B}}T_c^{\mathrm{AD}} \,=\,\frac{\omega_{\mathrm{log}}}{1.2}\,\exp\Biggl[-\,\frac{1.04\,(1+\lambda)}{\lambda(1-0.62\,\mu^{*})-\mu^{*}}\,\Biggr],
\end{equation}
where $\mu^*=0.125$. We also calculate $T^\text{E}_c$ from a numeric solution to the isotropic Eliashberg equations. The values of $T_c$ do not change significantly for a range of $\mu^*$ values from 0.06 to 0.18.

{\em Structure searches -} In the study of the lutetium-hydrogen binary system, we perform structure searches using the same Ephemeral Data-Derived Potential (EDDP)\,\cite{Pickard2022} as used by some of us in Ferreira and co-workers\,\cite{pickard-structure-prediction-arxiv}. To generate the initial binary structures, we remove hydrogen atoms in cubic LuH$_3$. Starting with a $2\times2\times2$ supercell of the pristine cubic LuH$_3$ structure, we enumerate all possible symmetrically inequivalent defect structures with hydrogen vacancies using the {\sc disorder} code\,\cite{code_disorder_1,code_disorder_2}. This results in 55,066 unique structures for compositions in the range LuH$_0$ to LuH$_3$, reducing the original count from 17,777,214 through symmetry considerations. These structures are then optimised using the potential. The low energy structures resulting from the machine learned potential relaxation are carried forward for subsequent DFT calculations.

We further use the \textit{ab initio} random structure searching (AIRSS) method\,\cite{Pickard2006,Pickard_AIRSS} to explore lutetium-hydrogen-nitrogen ternary compounds with cubic symmetry matching that reported in experimental X-ray diffraction data. We generate 1,000 cubic structures by randomly replacing all Lu and H sites with H, N, or a vacancy, which supplements the 200,000 randomly generated Lu-H-N structures sampled in previous work \cite{pickard-structure-prediction-arxiv}. We then relax these additional structures using DFT.

\section{Data availability}
The data that support the findings of this study are available within the paper and Supplementary Information. In particular, the dataset for dielectric functions and structure files can be found at \url{https://github.com/monserratlab/Lu_hydrides_colour}.

\section{Code availability}
The {\sc VASP} code used in this study is a commercial electronic structure modeling software, available from \url{https://www.vasp.at}.  The {\sc Quantum Espresso} code used in this research is open source: \url{https://www.quantum-espresso.org/}.

\section{References}
\bibliographystyle{naturemag}

\section{Acknowledgments}

\begin{acknowledgments}
S.-W.K. and B.M. are supported by a UKRI Future Leaders Fellowship [MR/V023926/1]. B.M. also acknowledges support from the Gianna Angelopoulos Programme for Science, Technology, and Innovation, and from the Winton Programme for the Physics of Sustainability. G.L.P. acknowledges funding from the Ministry of Research, Innovation, and Digitalisation within Program 1-Development of National Research and Development System, Subprogram 1.2-Institutional Performance-RDI Excellence Funding Projects, under contract no.\,10PFE/2021. The computational resources were provided by the Cambridge Tier-2 system operated by the University of Cambridge Research Computing Service and funded by EPSRC [EP/P020259/1], by the UK National Supercomputing Service ARCHER2, for which access was obtained via the UKCP consortium and funded by EPSRC [EP/X035891/1], and by the  SCARF cluster of the STFC Scientific Computing Department.

\textit{For the purpose of open access, the authors have applied a Creative Commons Attribution (CC BY) licence to any Author Accepted Manuscript version arising from this submission.}
\end{acknowledgments}

\section{Author contributions}

B.M. and S.-W.K. conceived the study and planned and supervised the research. S.-W.K. performed the DFT calculations, S.-W.K. and G.L.P. performed the DMFT calculations, L.J.C. and C.J.P. performed the structure searches, and all authors contributed to the analysis. B.M. and S.-W.K. wrote the manuscript with input from all authors.

\section{Competing interests}
The authors declare no competing interests.

\ifarXiv
    

\section*{Supplementary material}

**Supplementary Note 1. ELECTRONIC BAND STRUCTURE**

We show the electronic band structure and density of states of $Fm\bar{3}m$ LuH$_2$ in Fig. S1**a** calculated using semilocal DFT in the PBEsol approximation (DFT model in the main text). In agreement with earlier calculations, we find a metallic band structure with the Fermi level crossing highly dispersive bands that lead to an overall small density of states at the Fermi energy.

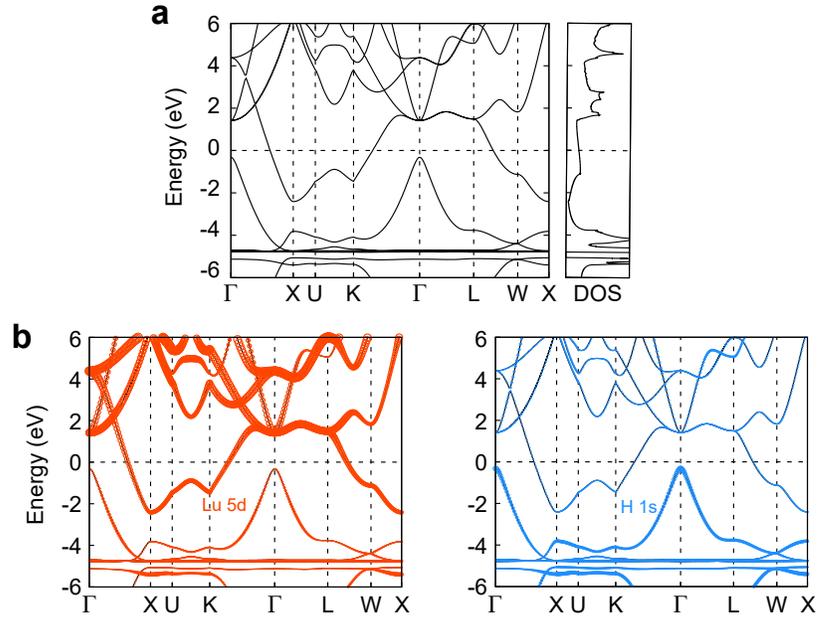

FIG. S1: **Electronic structure of pristine LuH$_2$. a.** Band structure and density of states of pristine LuH$_2$. **b.** Orbital-projected band structure of LuH$_2$. The band structure is projected onto lutetium $5d$ (left) and hydrogen $1s$ (right) orbitals and the size of the open circles is proportional to the projected weight.

We also show the electronic band structure projected on various lutetium and hydrogen atomic orbitals in Fig. S1**b**. The states derived from the lutetium $5d$ orbitals are highly dispersive.

Figure S2 depicts the evolution of the band structure of LuH$_2$ under pressure. Most bands only show rigid shifts with increasing pressure, apart from the bands near the X point in the Brillouin zone which exhibit a more significant re-organisation.

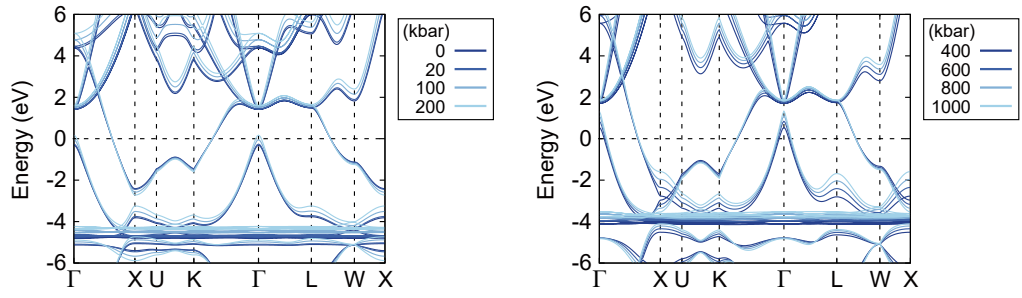

FIG. S2: **Pressure dependence of the electronic band structure of LuH$_2$.**

**Supplementary Note 2. ELECTRON CORRELATION**

Lutetium has a partially filled $5d$ shell which may lead to strong electronic correlations. In Fig. S3 we show the electronic band structure of $LuH_2$ calculated using DFT corrected with a Hubbard $U$ term (DFT+$U$ model in the main text). We show calculations using $U$ values in the range from 0 to 5 eV. The band structure changes with $U$ are small, which is consistent with the small changes reported in the main text for the reflectivity and colour. Similarly, we have tested the role of dynamical correlations using DFT augmented with dynamical mean field theory, and find negligible contributions.

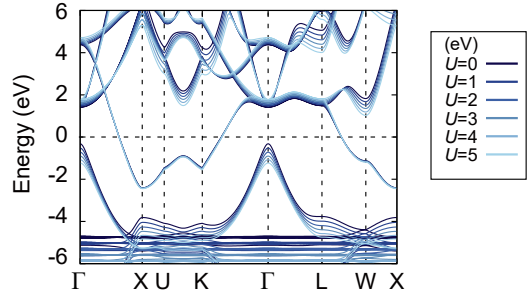

FIG. S3: **Dependence of the electronic band structure of $LuH_2$ on the Hubbard $U$ parameter.**

**Supplementary Note 3. ELECTRON-PHONON COUPLING**

Hydrogen is the lightest of all elements, and as such it exhibits significant quantum fluctuations which are responsible for the large electron-phonon coupling driving high temperature superconductivity in high pressure hydrides. In the main text, we present the reflectivity of $LuH_2$ under ambient conditions including the effects of electron-phonon interactions arising from quantum and thermal fluctuations. In this Supplementary Note we present additional details about the associated calculations.

As described in the main text, the dielectric function at temperature $T$ renormalized by electron-phonon coupling is given by [1, 2]:

$$\varepsilon_2(\omega; T) = \frac{1}{\mathcal{Z}} \sum_{\mathbf{s}} \langle \Phi_{\mathbf{s}}(\mathbf{u}) | \varepsilon_2(\omega; \mathbf{u}) | \Phi_{\mathbf{s}}(\mathbf{u}) \rangle e^{-E_{\mathbf{s}}/k_\mathrm{B}T}, \tag{1}$$

where $\mathcal{Z}$ is the partition function, $|\Phi_{\mathbf{s}}(\mathbf{u})\rangle$ is a harmonic eigenstate $\mathbf{s}$ of energy $E_{\mathbf{s}}$, $\mathbf{u} = \{u_{\mathbf{q}\nu}\}$ is a vector containing all atomic positions expressed in terms of normal mode amplitudes $u_{\mathbf{q}\nu}$, and $k_\mathrm{B}$ is Boltzmann's constant.

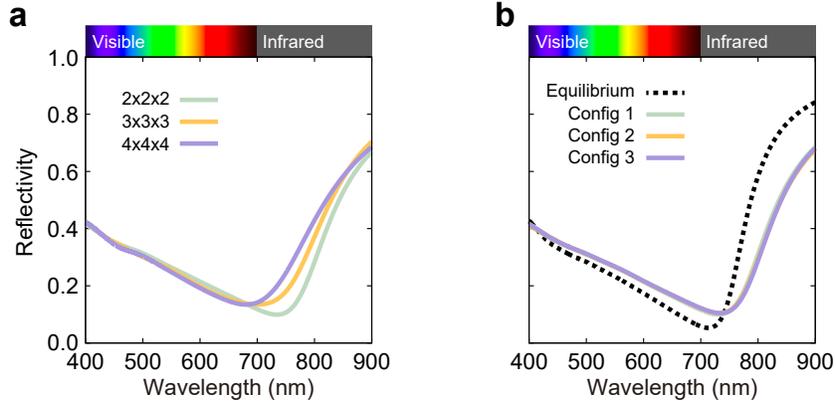

FIG. S4: **Convergence of electron-phonon coupling effects on the reflectivity of $LuH_2$.** **a.** Convergence with respect to supercell size. **b.** Convergence with respect to the number of stochastic configurations in the Monte Carlo integration.

We evaluate Eq. (1) using Monte Carlo integration accelerated with thermal lines [3]. We start from the phonon calculations described in Supplementary Note 5 and we then generate atomic configurations in which the atoms are distributed according to the harmonic nuclear

wave function in which every normal mode has an amplitude equal to:

$$u_{\mathbf{q}\nu} = \pm \left(\frac{1}{2\omega_{\mathbf{q}\nu}}[1 + 2n_{\mathrm{B}}(\omega_{\mathbf{q}\nu}, T)]\right)^{1/2}, \qquad (2)$$

where $n_{\mathrm{B}}(\omega, T)$ is the Bose-Einstein factor, and the sign of the amplitude is chosen stochastically. These calculations involve two convergence parameters which need to be tested. First, electron-phonon interactions need to be converged with respect to supercell size (or equivalently with respect to Brillouin zone grid size). Figure S4a shows the electron-phonon renormalised reflectivity calculated using increasingly large supercell sizes. There is some dependence on supercell size, but the overall reflectivity shape is relatively consistent between them. In the main text we use the results from the $4 \times 4 \times 4$ calculations. Second, in the Monte Carlo integration, the electron-phonon renormalised dielectric function is in principle obtained by averaging over stochastic configurations, where the number of configurations is a convergence parameter. We find that the electron-phonon renormalised reflectivity converges very rapidly with respect to the number of configurations included in the calculation: Fig. S4b shows three different configurations (compared to the equilibrium configuration) and the curves are indistinguishable. We use a single configuration for the results reported in the main text.

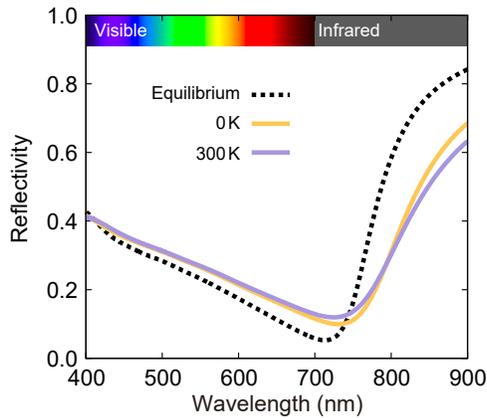

FIG. S5: **Quantum and thermal fluctuations effects on the electron-phonon coupling renormalised reflectivity of LuH$_2$.**

Our calculations including electron-phonon coupling are capable of describing the effects of both quantum fluctuations at $0\,\mathrm{K}$ and thermal fluctuations at finite temperature. To evaluate the relative contribution of quantum and thermal fluctuations, in Fig. S5 we compare

7

the reflectivity of LuH$_2$ obtained by fixing the nuclei at their equilibrium positions (no quantum nor thermal fluctuations), at 0 K (including only quantum fluctuations) and at 300 K (including quantum and thermal fluctuations). The difference between the equilibrium and 0 K curves quantifies the role of quantum fluctuations, which lead to a shift of the reflectivity minimum and to a decrease of the reflectivity in the infrared part of the spectrum, but the overall reflectivity shape does not change significantly. The difference between the 0 K and 300 K curves quantifies the role of thermal fluctuations, which make a significantly smaller contribution.

**Supplementary Note 4. LUTETIUM-HYDROGEN BINARY SYSTEM**

**4.1. Crystal structures of LuH$_2$ and LuH$_3$**

We have performed extensive structure searches for stoichiometries ranging from LuH$_0$ to LuH$_3$ and at multiple pressures. The results are summarised in the convex hull diagrams depicted in Fig. 2a in the main text. Here, we focus on the stoichiometric LuH$_2$ and LuH$_3$ as they are mostly relevant phases studied in the literature. We determine all competing structures in the pressure range 0-1000 kbar.

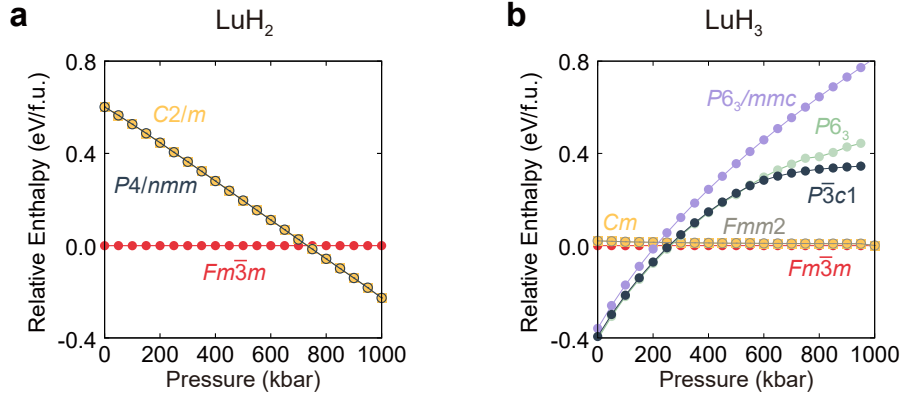

FIG. S6: **Relative enthalpy of LuH$_2$ and LuH$_3$.**

Figure S6 shows the relative enthalpy of the competing structures. For LuH$_2$, the ground state structure at 0 kbar has cubic $Fm\bar{3}m$ symmetry. A structure of $P4/nmm$ symmetry becomes more stable above 732 kbar. We note that there is a structure of space group $C2/m$ whose energy is almost degenerate with that of the $P4/nmm$ structure across the entire pressure range. In the main text, we report the pressure-driven reflectivity and colour changes of the $Fm\bar{3}m$ structure only. We report the reflectivity and colour of the $P4/nmm$ structure in Fig. S7.

For LuH$_3$, there are multiple competing phases in the pressure range 0-1000 kbar. The most important ones are the trigonal $P\bar{3}c1$ and the cubic $Fm\bar{3}m$ structures, as they are the most consistent with experimental observations. In the main text, we present the reflectivity and colour of $P\bar{3}c1$ at 0 and 100 kbar and $Fm\bar{3}m$ at 400 kbar according to the relative enthalpy. The full reflectivity and colour data of both structures over the entire pressure range of interest are presented in Fig. S8. However, we note that both of these structures are dynamically unstable at 0 kbar (see Fig. S15 for an example), and the most stable structure

at 0 kbar is the $P6_3$ structure, which is almost degenerate in energy with the $P\overline{3}c1$ structure. We note that the $P6_3$ structure is insulating up to a pressure of about 100 kbar. We also note that two structures with space groups $Cm$ and $Fmm2$ are almost degenerate in energy with the cubic $Fm\overline{3}m$ structure over the entire pressure range of interest (see Fig. S9 for the reflectivity and colour of the $Fmm2$ structure).

### 4.2. Reflectivity and colour of LuH$_2$ and LuH$_3$

Multiple structure searching works have predicted a range of stable and metastable structures in the binary Lu-H system that may play a role in the reported superconductivity, and we report our own structure searching results in the main text. Here, we present the reflectivities and colours of the most important phases of LuH$_2$ (Fig. S7) and LuH$_3$ (Fig. S8).

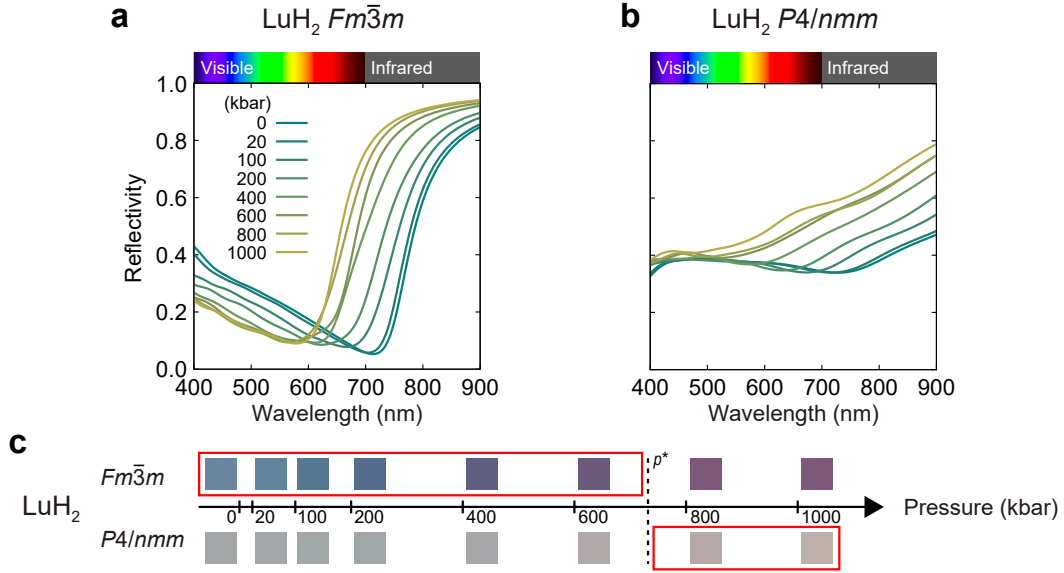

FIG. S7: **Reflectivity and colour of LuH$_2$.**

Additionally, we show the calculated reflectivities and colours for the LuH$_3$ structures with space groups $Fmm2$ and $P6_3/mmc$ in Fig. S9. Both structures exhibit a rather featureless reflectivity which leads to a light grey colour for $Fmm2$ LuH$_3$ and to a dark grey colour for $P6_3/mmc$ LuH$_3$. We have also calculated the reflectivity of the $C2/m$ LuH$_2$ structure (not shown), and it is very similar to that of the $P4/nmm$ structure, to which it is almost degenerate in enthalpy (Fig. S6).

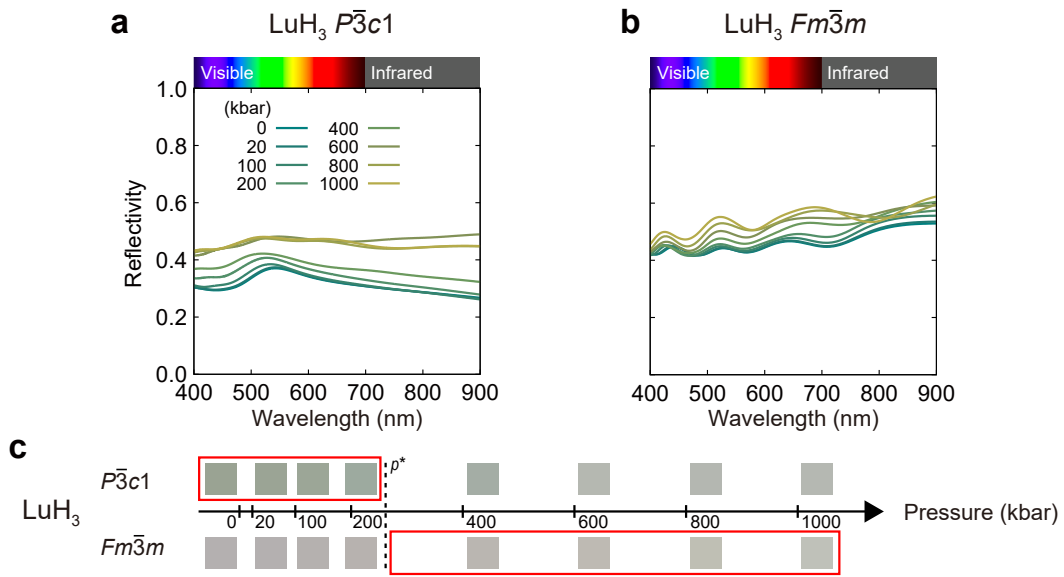

FIG. S8: **Reflectivity and colour of LuH$_3$.**

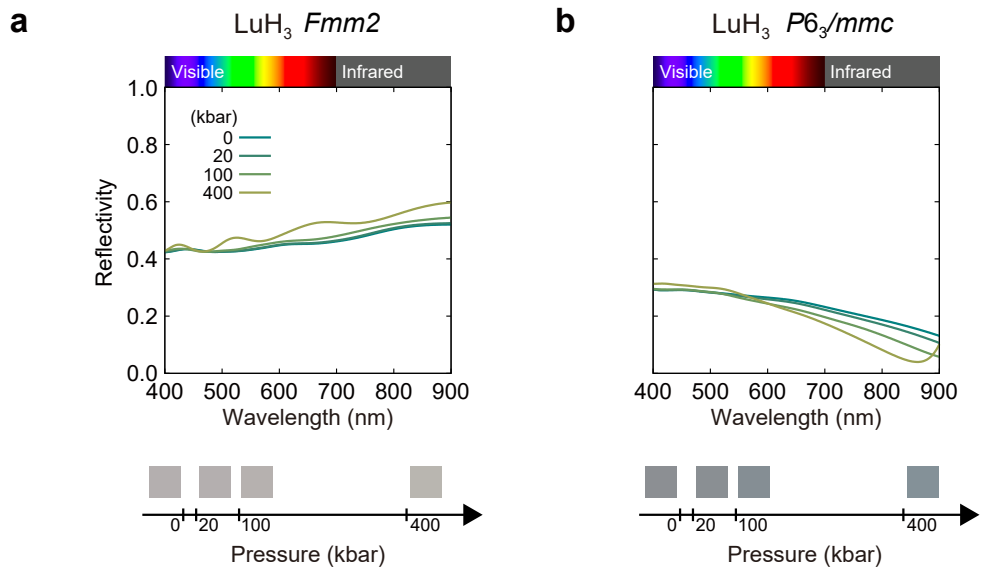

FIG. S9: **Reflectivity and colours of $Fmm2$ and $P6_3/mmc$ LuH$_3$ structures.**

11

### 4.3. Reflectivity and colour of nitrogen doped cubic $LuH_2$ and $LuH_3$

We establish hydrogen-deficient $LuH_2$ as the dominant phase responsible for the experimentally observed colour changes. However, multiple experimental reports include a small amount of nitrogen doping, and in this subsection we provide an overview of the interplay between nitrogen doping and colour in $LuH_{2-\delta}$ and $LuH_{3-\delta}$.

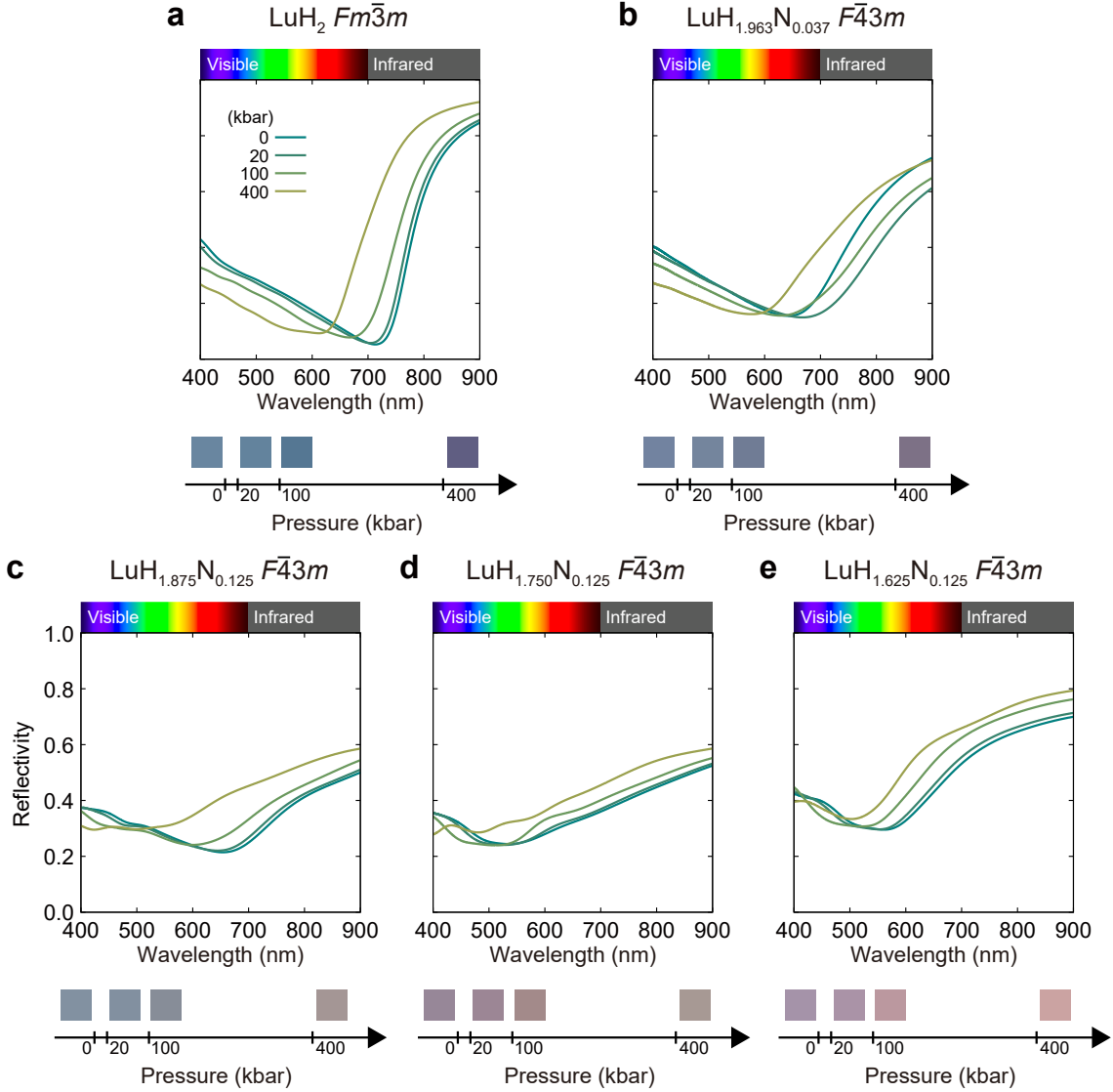

FIG. S10: **Reflectivity and colour of $LuH_{2-\delta}N_\epsilon$ structures.** $LuH_{1.963}N_{0.037}$ is simulated as a single hydrogen vacancy and a single nitrogen addition in a $3 \times 3 \times 3$ supercell. Other structures are simulated in a $2 \times 2 \times 2$ supercell.

We show the pressure-driven changes in the reflectivity and colour of nitrogen-doped

12

LuH$_{2-\delta}$ in Fig. S10. The effect of nitrogen doping is to flatten the reflectivity curves obtained with LuH$_{2-\delta}$, such that the resulting colours become more grey. As a result, we conclude that nitrogen doping also affects the reflectivity and colour of LuH$_2$, but the changes are secondary compared to those driven by hydrogen vacancies.

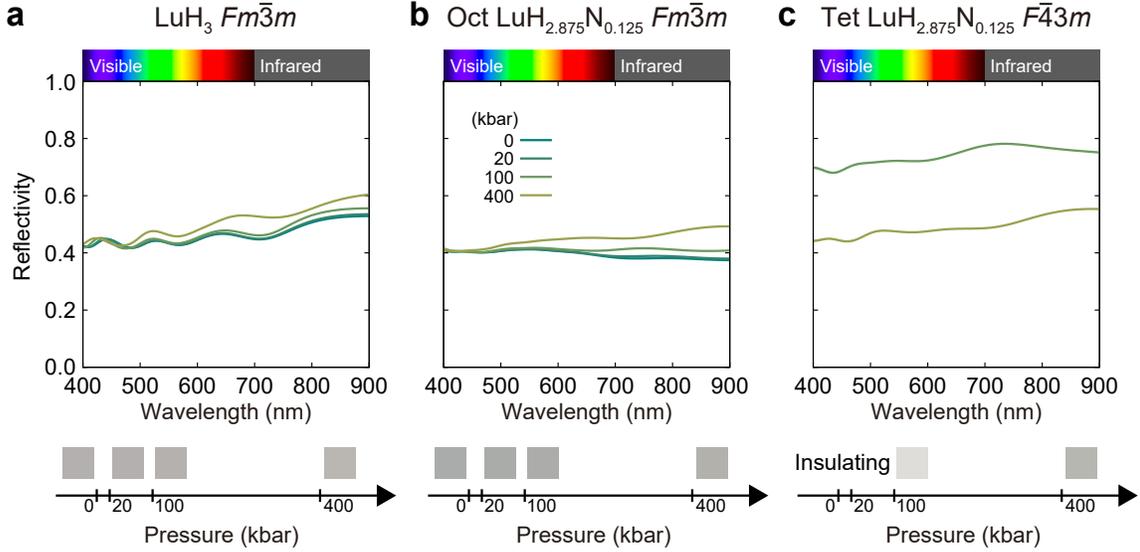

FIG. S11: **Reflectivity and colour of LuH$_{3-\delta}$N$_\epsilon$ structures.** The single hydrogen vacancy in LuH$_3$ is simulated in a $2 \times 2 \times 2$ supercell. In **b** and **c**, Oct and Tet denote the octahedral and tetrahedral sites for the hydrogen vacancy, respectively. In **c**, the structure is insulating at low pressures.

For completeness, we also show the pressure-driven changes in the reflectivity and colour of nitrogen-doped LuH$_{3-\delta}$ in Fig. S11. Nitrogen doping has a small effect in the reflectivities, which remain relatively flat leading to overall grey colours.

13

# Supplementary Note 5. LUTETIUM-HYDROGEN-NITROGEN TERNARY SYSTEM

## 5.1. Crystal structures

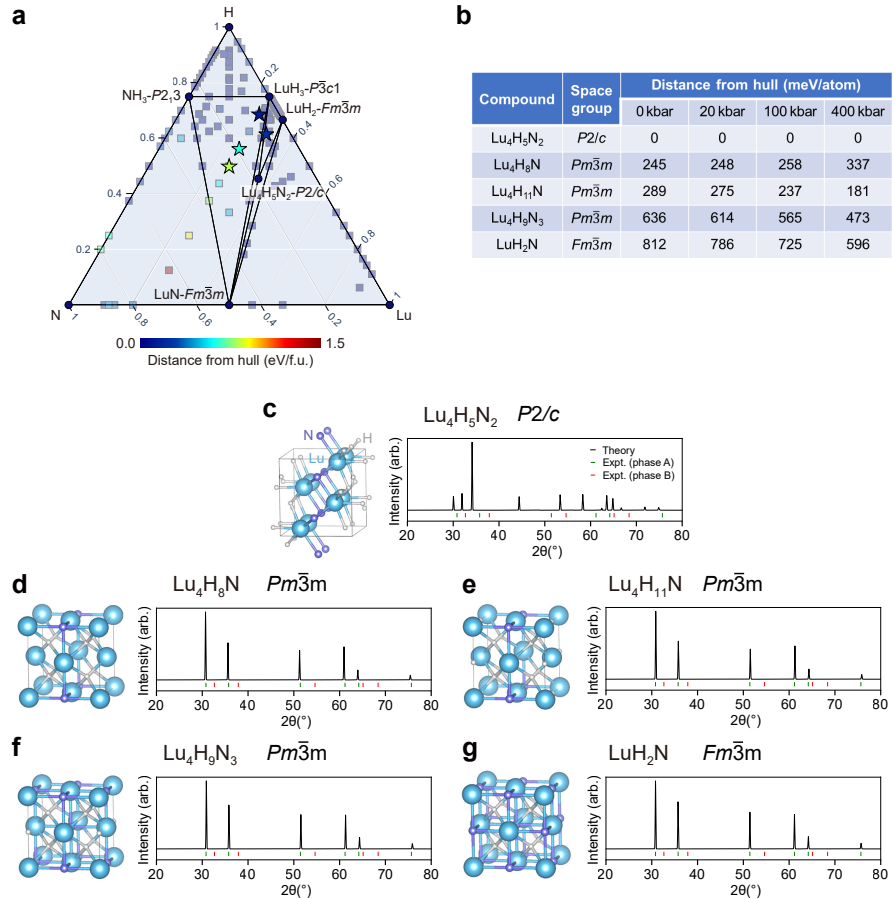

FIG. S12: **Convex hull diagram, energetics, and atomic structures of the lutetium-hydrogen-nitrogen ternary system. a.** Convex hull diagram for the Lu-H-N ternary system at ambient pressure. Dark blue circles indicate thermodynamically stable structures and metastable structures are shown as squares and stars. Stars indicate the cubic structures whose simulated X-ray diffraction (XRD) pattern is consistent with the experimental report [4], as shown in **d-g**. The colour scale represents the energy distance from the convex hull. **b.** Based on the calculated distance from the hull, we present the energetics of the five best Lu-H-N compounds including thermodynamically stable $Lu_4H_5N_2$ as a function of pressure. **c-g.** Atomic structures and XRD simluation results for **c** $Lu_4H_5N_2$, **d** $Lu_4H_8N$, **e** $Lu_4H_{11}N$, **f** $Lu_4H_9N_3$, and **g** $LuH_2N$. Except for the $Lu_4H_5N_2$ compound with $P2/c$ symmetry, Lu atoms form a fcc lattice. For the XRD similulations, experimental XRD data are displayed as green and red ticks, which are assigned as two different phases in Ref. [4].

We perform extensive crystal structure search in the full Lu-H-N ternary space (Fig. S12). We highlight that our structure searches [5] are the only ones of all those published that identify a stable ternary compound $P2/c$ $Lu_4H_5N_2$ at ambient pressure. We have selected multiple ternary Lu-H-N compounds $Pm\bar{3}m$ $Lu_4H_8N$, $Pm\bar{3}m$ $Lu_4H_{11}N$, $Pm\bar{3}m$ $Lu_4H_9N_3$, and $Fm\bar{3}m$ $LuH_2N$ that are not on the convex hull but whose simulated X-ray diffraction data is consistent with experimental reports. The calculated colours for these compounds are shown in Fig. S13.

### 5.2. Reflectivity and colour

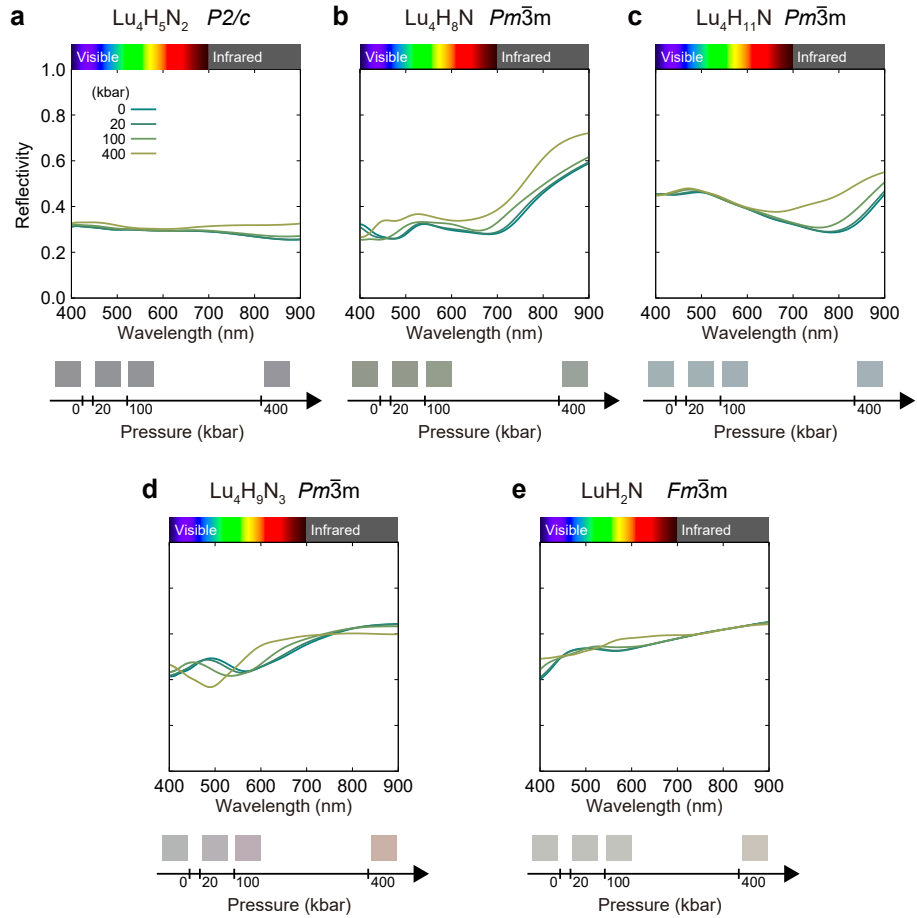

FIG. S13: **Pressure dependence of reflectivity and colour of the lutetium-hydrogen-nitrogen ternary system.**

Multiple structure searching works have predicted a stable and multiple metastable struc-

15

tures in the ternary Lu-H-N system that may play a role in the reported superconductivity. $Lu_4H_5N_2$ is the only structure predicted to be thermodynamically stable in the Lu-H-N ternary system, and we show its reflectivity and colour as a function of pressure in Fig. S13**a**. Its reflectivity in the visible range is small and constant, leading to an overall dark grey colour. We also display the reflectivities and colours for four additional metastable structures of stoichiometries $Lu_4H_8N$, $Lu_4H_{11}N$, $Lu_4H_9N_3$, and $LuH_2N$ in Figs. S13**b-e**. These structures are metastable but all have cubic symmetry, which has been identified experimentally as the relevant symmetry. Their reflectivities are relatively constant across the visible range of the spectrum, leading to overall grey colours with a slight blue tone in $Lu_4H_{11}N$ at ambient pressure.

**Supplementary Note 6. PHONON DISPERSION**

Regarding the dynamical stability of the compounds, we have confirmed that the pure $LuH_2$ and hydrogen deficient $LuH_{2-\delta}$ structures, depicted in photorealistic rendering in Fig. 3 in the main text, exhibit dynamic stability at the harmonic level, as displayed in the phonon dispersion in Fig. S14. The phonon band structures show a large gap separating the low-frequency acoustic modes from the high frequency optical modes. This large gap is expected in binary compounds with a large mass difference between the elemental components, with hydrogen having a relative atomic mass of 1.008 and lutetium of 174.967.

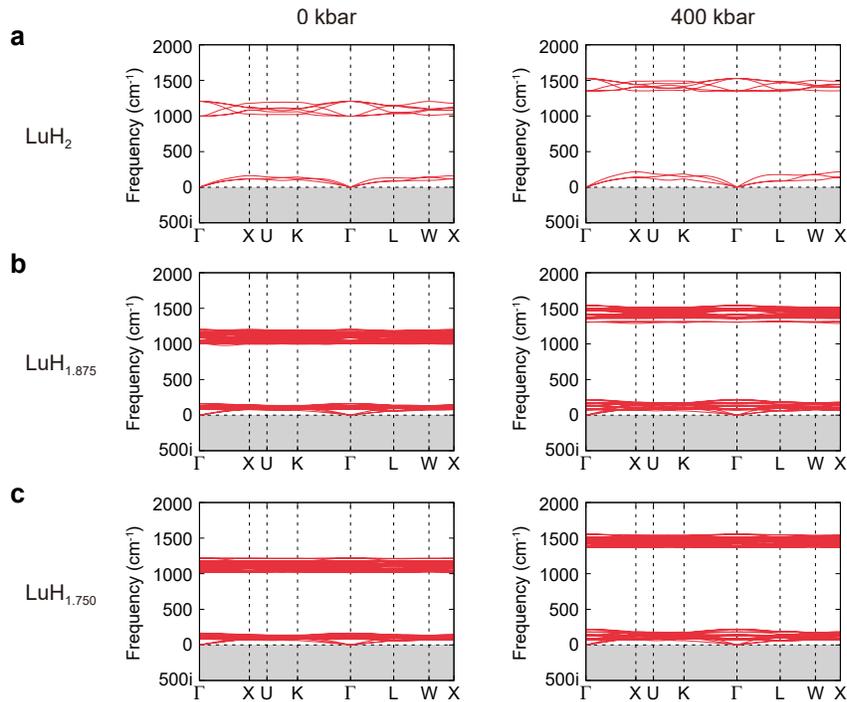

FIG. S14: **Phonon dispersion of pure and hydrogen-deficient lutetium dihydrides at** $0$ **and** $400$ **kbar.**

We have also confirmed the dynamical stability of multiple other structures (e.g. LuH in Fig. S15a). However, we note that cubic $LuH_3$, proposed as the phase responsible for superconductivity in the original Nature paper by Dias and co-workers, is dynamically unstable at the harmonic level (Fig. S15b). This has also been reported by other works.

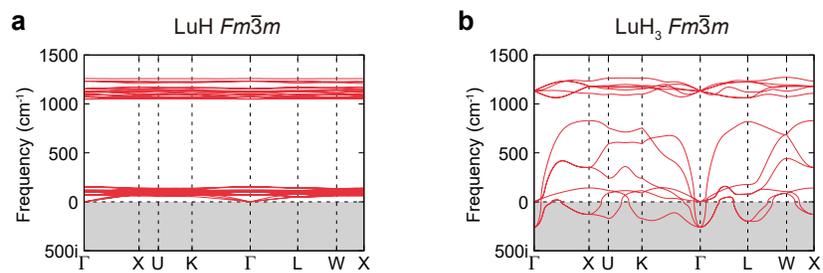

FIG. S15: **Phonon dispersion of lutetitum monohydride and trihydride.** Results are shown for **a** LuH in the $Fm\bar{3}m$ structure and **b** LuH$_3$ in the $Fm\bar{3}m$ structures.

**Supplementary Note 7. PHONON-MEDIATED SUPERCONDUCTIVITY**

There have been several reports of phonon-mediated superconductivity calculations in a range of stable and metastable compounds of the Lu-H-N system. Interestingly, none of these calculations find superconducting critical temperatures near room temperature. Experimentally, the superconducting phase is claimed to only exist when the sample exhibits a pink colour. Given our discovery that the pink phase of lutetium hydride only exists in hydrogen-deficient $LuH_2$, we estimate the phonon-mediated superconducting critical temperature of $LuH_{1.875}$ and $LuH_{1.750}$ as a function of pressure. Figure S16 shows the isotropic Eliashberg function $\alpha^2 F(\omega)$ including the pink colour phases of $LuH_{1.875}$ at 400 kbar and $LuH_{1.750}$ at 20 kbar. The corresponding superconducting critical temperatures for $LuH_{1.875}$ ($LuH_{1.750}$) at 20, 100, and 400 kbar are 0.19 K (0.04 K), 0.16 K (0.03 K), and 0.33 K (0.06 K), respectively.

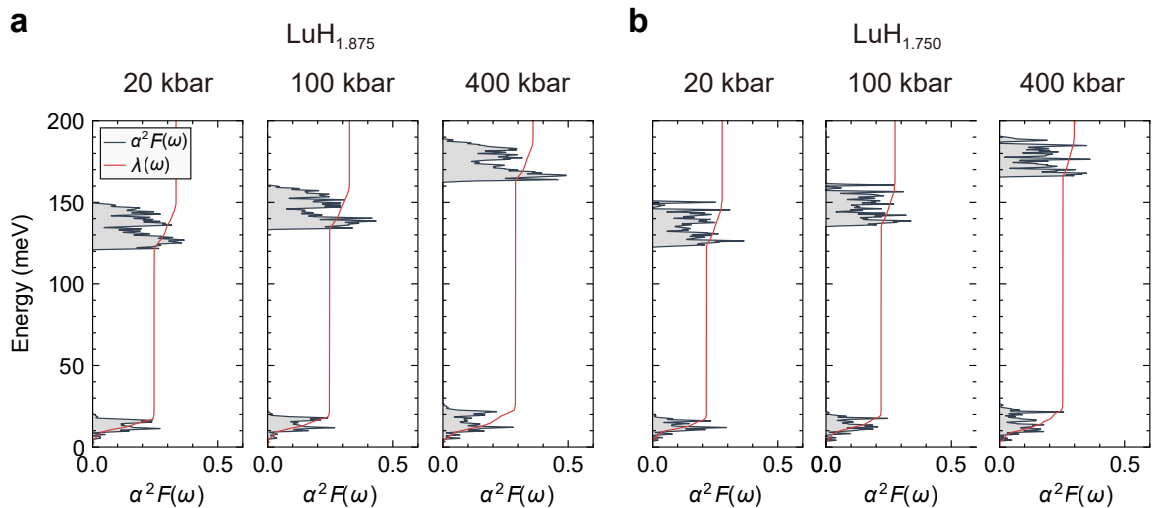

FIG. S16: **Isotropic Eliashberg function of the hydrogen-deficient $LuH_{1.875}$ and $LuH_{1.750}$ at multiple pressures.**

**Supplementary Note 8. DETAILS OF PHOTORELIASTIC RENDERING**

**8.1. Benchmark tests on Lu and LuH$_2$**

We have described the computational methodology used to calculate colour in Methods, which was originally developed and tested by Prandini and co-workers [6]. They applied the methodology to a wide variety of metallic systems, including 18 elemental metals (e.g. Au, Cu, Al), intermetallic compounds (e.g. AuAl$_2$, PdIn), and solid solutions (e.g. Au-Ag solid solutions), in all cases demonstrating remarkable agreement between theory and experiment. Therefore, the computational method we use has been thoroughly validated for many compounds. As an additional validation, in our work we have tested the calculation of the colour of ambient conditions LuH$_2$ and Lu, and again our results are in good agreement with experiment (Fig. S17).

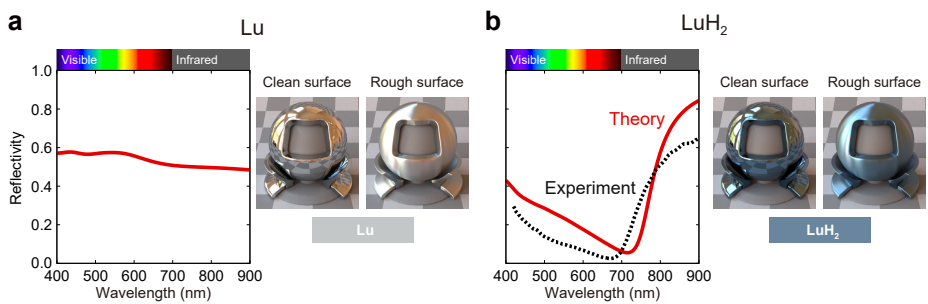

FIG. S17: **Reflectivity and colour of Lu and LuH$_2$.** Here, we present two photorealistic rendering images for each system with different surface roughness conditions (clean or rough surface) as implemented in the MITSUBA 3 renderer [7]. The photorealistic rendering of the Lu metal shows a silvery white colour, in good agreement with experiment [8].

**8.2. Surface roughness effect**

Figure S18 illustrates the blue colour of LuH$_2$ using different models for the surface roughness, which shows the type of change that one may expect between different samples. We find that the surface effect does not change the colour. We highlight that the key quantity to describe the optical response of materials is the dielectric function, and we include all calculated dielectric functions as Supplementary Data to enable readers to evaluate the optical response for distinct setups.

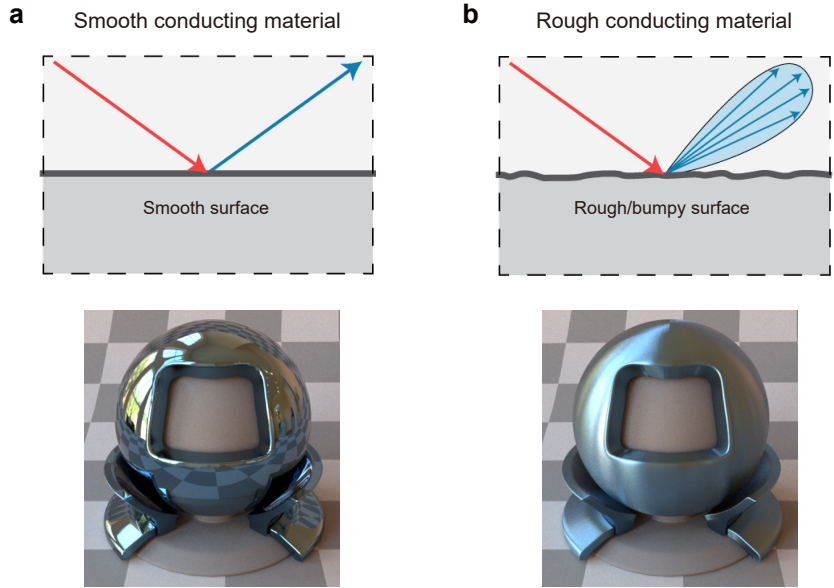

FIG. S18: **Surface roughness effect.** Photorealistic rendering images for LuH$_2$ with different surface roughness conditions (clean or rough surface) as implemented in the MITSUBA 3 renderer [7]. The images in upper panels are taken from this webpage on the MITSUBA 3 renderer website. For **b**, we use roughness values of $\alpha_u = 0.05$ and $\alpha_v = 0.3$.

### 8.3. Choice of exchange-correlation functional

We note that the brightness of the colour can change if we use different exchange-correlation functionals for the DFT calculations. For example, Figure S19 shows DFT+$U$ results compared with DFT, with the former exhibiting brighter orange colours and a transition to a red phase in pure LuH$_2$. Despite these small change depending on the calculation details, the overall conclusions of our work remain unchanged.

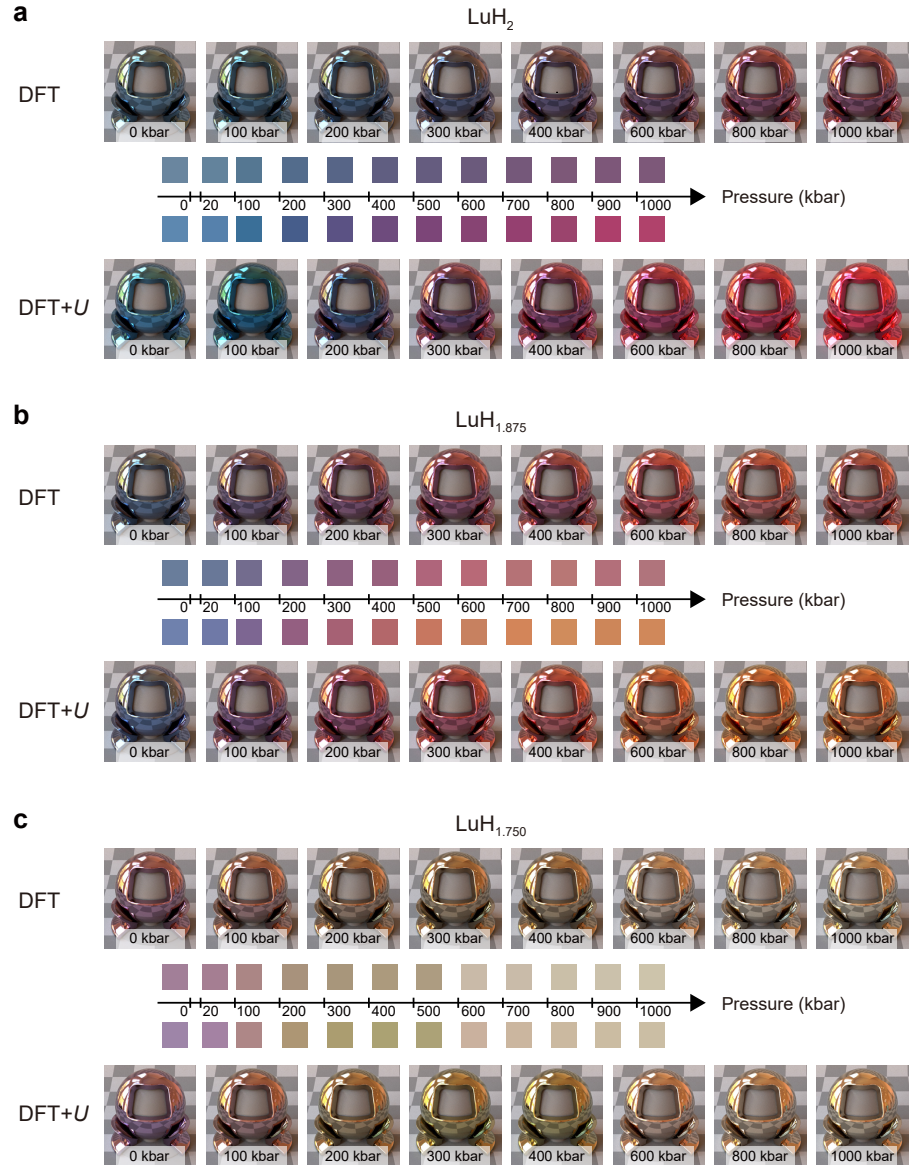

FIG. S19: **Comparison between DFT and DFT+$U$ colour results. a-c.** Colour and photorealistic rendering of **a** $LuH_2$, **b** $LuH_{1.875}$ and **c** $LuH_{1.750}$ as a function of pressure. We use $U = 3\,\text{eV}$ on Lu $d$ orbitals.

**Supplementary Note 9. CONVERGENCE TESTS ON REFLECTIVITY**

We have tested various convergence parameters on the calculation of reflectivity, and found that an energy cutoff of $400\,\text{eV}$ and a **k**-point grid size of $40 \times 40 \times 40$ are fully converged. Convergence test details are presented in Fig. S20.

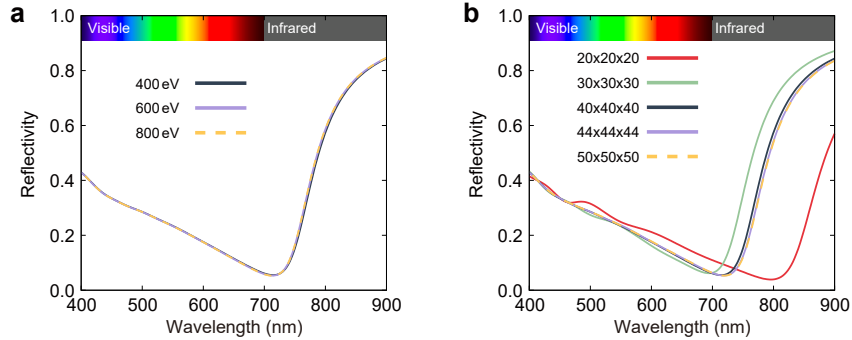

FIG. S20: **Convergence tests on reflectivity.**


\fi

\end{document}